\documentclass[10pt,twocolumn,aps,pra,notitlepage,showpacs,floatfix,amsmath]{revtex4-1}
\usepackage{graphicx}
\usepackage{amssymb}
\usepackage{epstopdf}
\usepackage{amsmath,amsfonts,latexsym}
\newcommand{\vect}[1]{\mathbf{#1}}
\newcommand{\be}{\begin{equation}}
\newcommand{\ee}{\end{equation}}
\newcommand{\bea}{\begin{eqnarray}}
\newcommand{\eea}{\end{eqnarray}}
\newcommand{\ba}{\begin{align}}
\newcommand{\ea}{\end{align}}

\newcommand{\NPP}{{$\text{N}_{\text{PP}}$}}
\newcommand{\LOg}{{$\text{LO}_\text{g}$}}
\newcommand{\LOgf}{\text{LO}_{\text{g}}}
\newcommand{\fourvec}[4]{\left[\begin{array}{cc} #1 & #2 \\ #3 & #4 \end{array} \right]}
\newcommand{\twovec}[3][2]{\left[\begin{array}{c} #2 \\ #3 \end{array} \right]}

\def\pra{\ref@jnl{Phys.~Rev.~A}}
\def\prb{\ref@jnl{Phys.~Rev.~B}}
\def\prc{\ref@jnl{Phys.~Rev.~C}}
\def\prd{\ref@jnl{Phys.~Rev.~D}}
\def\pre{\ref@jnl{Phys.~Rev.~E}}
\def\prl{\ref@jnl{Phys.~Rev.~Lett.}}

\usepackage{amsmath,amsthm,amsfonts,amssymb,mathrsfs,mathtools}
\usepackage{enumerate}
\usepackage{color}
\usepackage[bookmarks=false,hyperfootnotes=false]{hyperref}  
\hypersetup{
			colorlinks=true,
			linkcolor=black,
			urlcolor=blue,
			anchorcolor=black,
			citecolor=blue,
			urlcolor=black,
			pdftitle={Larkin-Ovchinnikov superfluidity in a two-dimensional imbalanced atomic Fermi gas},
			pdfauthor={Umberto Toniolo}
}

\renewcommand{\theta}{\vartheta}

\renewcommand{\epsilon}{\varepsilon}

\DeclareMathOperator{\sgn}{sgn}
\DeclareMathOperator{\im}{i \hspace{-0.2mm}}

\newcommand{\defeq}{\coloneqq}

\newcommand{\NN}{\mathbb{N}}	
\newcommand{\ZZ}{\mathbb{Z}}	

\newcommand{\conj}[1]{\overline{#1}}

\begin{document}

\preprint{APS/123-QED}

\title{Larkin-Ovchinnikov superfluidity in a two-dimensional imbalanced atomic Fermi gas}

\author{Umberto Toniolo}
\email{utoniolo@swin.edu.au}

\author{Brendan Mulkerin}
\author{Xia-Ji Liu}
\author{Hui Hu}

\affiliation{Centre for Quantum and Optical Science, Swinburne University of Technology, Hawthorn 3122 VIC, Australia}

\date{\today} 

\begin{abstract}
We present an extensive study of two-dimensional Larkin-Ovchinnikov (LO) superfluidity in a spin-imbalanced two-component atomic Fermi gas. In the context of Fulde-Ferrell-Larkin-Ovchinnikov (FFLO) phase, we explore a wide and generic class of pairing gap functions with explicit spatial dependency. The mean-field theory of such phases is applied through the Bogoliubov-de Gennes equations in which
the pairing gap can be determined self-consistently. In order to systematically explore the configuration space we consider both the canonical and grand canonical ensembles where we control the polarization or chemical potentials of the system, respectively. The mean-field calculations enable us to understand the nature of the phase transitions in the fully paired Bardeen-Cooper-Schrieffer (BCS) state, exotic LO phase, and partially polarized free Fermi gas (\NPP). The order of the phase transitions has been examined and in particular we find a weak first-order phase transition between the exotic inhomogeneous LO phase and the BCS phase. In comparison to the three-dimensional case, where the phase diagram is dominated by a generic separation phase, we predict a broader parameter space for the spatially inhomogeneous LO phase. By computing the superfluid density of the LO phase at different polarization, we show how the superfluidity of the system is suppressed with increasing spin polarization.
\end{abstract}

\pacs{03.75Ss, 03.75.Hh, 74.20.-z}

\maketitle

\section{\label{sec:level1}Introduction}

Superfluidity in fermionic systems has been experimentally and theoretically explored since the seminal experiments performed with $^3$He in the early 1970's~\cite{PhysRevLett.28.885,PhysRevLett.29.920}. Here it was found that the critical temperature was roughly a thousand times smaller than the bosonic $^4$He superfluid transition temperature. The statistical properties of the fermonic system led theorists to consider an analogue of the successful Cooper pairs ansatz used in the Bardeen-Cooper-Schrieffer (BCS) theory, describing the superfluid state by a superconducting theory involving particles without electric charge~\cite{PhysRevLett.29.1227}. Although successful in describing the low temperature behavior of the $^3$He system, the BCS theory was an incomplete treatment for more complicated systems, such as the case of  imbalanced Fermi mixtures. A finite polarization, due to Pauli paramagnetism, is strongly opposed to the BCS predicted ground state.
  
The first theoretical attempt to explore polarized fermionic systems was independently performed by Fulde and Ferrell (FF)~\cite{FuldeFerrell64} and by Larkin and Ovchinnikov (LO)~\cite{LarkinOvchinnikov65} and the resulting phase has since then been referred to as the FFLO phase. Starting from the Cooper solution of a fermionic pair that experiences an attractive potential over a Fermi surface, they considered the case where the pair could carry a finite center-of-mass momentum. This situation in the balanced case is energetically costly. It was pointed out by Cooper that pairs with finite center-of-mass momentum are energetically unfavorable due to the increase of internal energy from the pair-drift velocity~\cite{PhysRev.108.1175}. In the case of an imbalanced system, where the Fermi surfaces of different components do not fully overlap, however, the most favorable energetic configuration of the ground state has a spatially inhomogeneous order parameter, compared to the BCS solution of a uniform order parameter. In an imbalanced system, Fulde and Ferrel proposed an ansatz for the order parameter of the form $\Delta(\vect{x})=\Delta_0\exp(\im\vect{Q}\cdot\vect{x})$, where $\Delta_0$ is a constant parameter and $\vect{Q}$ plays the role of the non-vanishing pair center-of-mass momentum~\cite{FuldeFerrell64}. Larkin and Ovchinnikov proposed the sum of two opposite plane-waves with the same momentum, $\Delta(\vect{x})=\Delta_0\cos(\vect{Q}\cdot\vect{x})$~\cite{LarkinOvchinnikov65}. 

Understanding these exotic inhomogenous superfluid phases has attracted considerable experimental and theoretical attention over the past fifty years in different branches of physics. The study of the FFLO phase has been of great interest to the condensed matter community, where  the FFLO could be a candidate phase in heavy-fermion superconductors like CeCoIn$_5$ at large in-plane Zeeman fields~\cite{Steglich1996,Tachiki1996,Bianchi2003}. The observation of anisotropic conductivity in organic superconducting compounds indicates that these materials may represent an interesting system to accommodate this inhomogeneous order parameter phase~\cite{Lortz2007,Bergk2010}. The FFLO state is of interest in quantum chromodynamics where it may be favoured at low temperature and high density \cite{Casal2004}. However, so far, conclusive signatures of  a spatially inhomogeneous superfluid phase have yet to be clearly demonstrated. 

Over the past decade, ultracold Fermi gases have offered a unique environment to explore population imbalanced fermionic systems~\cite{Zwierlein2006,PhysRevLett.97.030401,Sheehy2006,Radzihovsky2010}. The controllability of ultracold experiments allows the tuning of interactions between fermions through Feshbach resonances to access a broad range of different interaction regimes~\cite{Inouye1998,Chin2010}. The combination of available techniques, such as optical lattices~\cite{Greiner2002,olattice2}, radio-frequency driven (spin) population imbalance~\cite{PhysRevA.78.033614}, and exotic interactions such as spin-orbit coupling \cite{Liu2013,Wu2013}, has opened up new methods to probe the existence of any novel phases that are strongly believed to occur at low temperatures. %
However, the existence of the FFLO phase in cold-atom systems is yet to be confirmed. The lack of experimental observation of the FFLO phase in three dimensions can be understood from the fragile nature and small region of the phase space where the FFLO is energetically favorable~\cite{PhysRevB.71.214504,PhysRevLett.95.117003,PhysRevA.83.063621}. 

A promising method to study imbalanced Fermi gases and the FFLO phase is through evolving the system in reduced dimensions. Anisotropy-related effects like exotic inhomogenous superfluid phases can be enhanced and stabilized by Fermi surface ``nesting" in low dimensional systems~\cite{PhysRevLett.99.120403,Koponen2008,Wolak2012}. A full one-dimensional (1D) configuration~\cite{Liu20071D} and a suitable experimental setup has been investigated~\cite{Liao2010}, providing new interesting hints for the FFLO-phase observation and detection. In two dimensional (2D) systems the problem has been theoretically analyzed, by testing the FF ansatz~\cite{Conduit2008,Sheehy2015} or LO ansatz with linearized single-particle dispersion relation~\cite{PhysRevB.30.122}, by considering square lattices within both FF and LO theories ~\cite{Baarsma2016},  and by investigating beyond mean-field effects with an FF order parameter~\cite{Torma1}. It is worth noting that, recently, there has been rapid progress in experimentally probing 2D Fermi gases~\cite{Murthy2015,Ong2015,Fenech2016,Dyke2016}. We anticipate that these advancements will be of great importance in finding the FFLO phase. 

\begin{figure}
\begin{centering}
\includegraphics[width=0.48\textwidth]{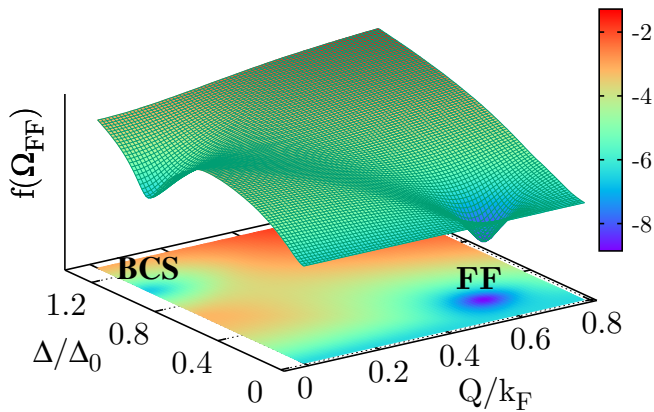} 
\protect\protect\caption{\label{fig:tp_ff}(color online). Thermodynamic potential per particle for the FF ansatz in 2D.  The $z$ axis is re-scaled by using a logarithmic function $f(z)=\log(1+\alpha z/(2\pi \epsilon_F))$, with $\alpha=9.916$.}
\par\end{centering}
\end{figure}

Theoretical results suggest that the FF ansatz is not the most energetically favorable choice and that the LO ansatz describes the preferred ground state~\cite{Radzihovsky2010}. In this work, we will build upon this idea and extensively explore the LO-type ansatz in 2D imbalanced Fermi gases through the Bogoliubov-de Gennes (BdG) mean-field theory, and determine the phase space where the LO-type phase is energetically favorable.  


This work is structured as follows. In Sec.~\ref{sec:fflo_family} we describe the FFLO family and set the appropriate notations of the ansatz of the order parameter to be used in this paper. The model is described in Sec.~\ref{sec:model}, including the relevant scattering theory, BdG formulation and computational implementation of the model. Section~\ref{sec:results} is devoted to the results where we discuss the canonical ensemble first and then the grand canonical ensemble. Section~\ref{sec:sf} contains the results of the superfluid density of the LO phase. Finally, in Sec.~\ref{sec:conclusions} we present our conclusions.

\section{\label{sec:fflo_family}The FFLO family}

For the sake of clarification we provide a summary of the definitions of the main cases in the FFLO family. The most generic order parameter for an imbalanced Fermi gas in a box with sides $L$ and volume $V=L^\nu$, for arbitrary %
dimension $\nu>1$ that satisfies periodic boundary conditions,  %
is a Fourier transform over the discretized modes of the box,   
\be\label{eq:fflo_family}
\Delta_{\text{FFLO}}(\vect{x})=\sum_{\vect{Q}}\Delta_{\vect{Q}}\exp(\im\vect{Q}\cdot\vect{x}),
\ee
with $\vect{Q}\in\ZZ^\nu$. Within this family many further approximations and simplifications may be made, %
for example, we can choose a single vector $\vect{Q}$ and obtain the FF ansatz for the order parameter, 
\be
\Delta_{\text{FF}}(\vect{x})=\Delta_{\vect{Q}}\exp\left(\im\vect{Q}\cdot\vect{x}\right).
\ee
We show in Fig.1 the thermodynamic potential per particle for the FF ansatz in 2D for chemical potential, $\mu=\varepsilon_F$, chemical potential difference, $\delta\mu=0.65\varepsilon_F$, and interaction strength, $\varepsilon_B=0.35\varepsilon_F$, as a function of the pair momentum $Q=|\mathbf{Q}|$. The order parameter is rescaled with the 2D BCS order parameter, $\Delta_0 = \sqrt{\varepsilon_B(\varepsilon_B+2\mu)}$ in the balanced limit.
The appearance
of the FF minimum in Fig.~\ref{fig:tp_ff}, at $Q \sim k_F$, indicates a first order phase transition from the BCS minimum.  The transition
from FF to the partially polarized normal state is second order, at which the FF minimum merges with the free Fermi gas line ($\Delta=0$). 
These results are found 
in the thermodynamic limit where the %
saddle point approximation of the thermodynamic potential can be analytically written by integrating %
the fermionic path integrals after a Hubbard-Stratonovich transformation. This case %
has been extensively studied and used to %
understand the generic features of the FFLO phase in 2D \cite{Conduit2008,Sheehy2015}. %
The FF ansatz, however, is not the most energetically favourable ground state. It has been shown that, for general dimensions, a sum over $\vect{Q}$ momenta is favourable \cite{Yoshida2007,bulgac2008}. 

If we consider the summation in Eq.~\eqref{eq:fflo_family}  of momenta $\vect{Q}$ and $-\vect{Q}$,  we find a standing wave, the LO anstaz,%
\be\label{eq:lo_ansatz}
\Delta_{\text{LO}}(\vect{x})=\Delta_{\vect{Q}}\cos(\vect{Q}\cdot\vect{x}).
\ee
The LO ansatz suffers from the loss of an analytic formula for the thermodynamic potential as the fermionic fields %
cannot be integrated out analytically from the partition function. To this end, some efforts to %
obtain a truncated approximation of the thermodynamic potential in 3D have provided %
interesting qualitative results~\cite{PhysRevA.87.063612}.

The LO ansatz in Eq.~\eqref{eq:lo_ansatz} can be further simplified by requiring the order parameter to %
depend on a single direction. It has been shown in 2D and 3D that the spontaneous breaking of symmetry due to %
the LO transition essentially occurs in a random direction only, and in 3D it occurs in a very narrow region %
of the phase diagram \cite{Radzihovsky2010}. The cost of the diagonalization of the Hamiltonian can be reduced in complexity and %
a complete study becomes tractable. Specifically, we have for a given momentum $\mathbf{Q}$ along the $x$-axis, the pairing order parameter, 
\be\label{eq:lo_one_ansatz}
\Delta_{\text{LO}_{\text{1}}}(\vect{x})=\Delta_{Q}\cos(Qx),
\ee
and we refer to this as the LO$_1$ pairing order parameter. This approach allows us to model the order parameter as a much more detailed function as long as we keep the phase transition symmetry breaking in one direction. In particular, %
we are able to study a generalization of the LO$_1$ phase that will be called \LOg~through %
the whole work,
\be\label{eq:lo_g_ansatz}
\Delta_{\LOgf}(\vect{x})=\sum_{Q}\Delta_{Q}\cos(Qx),
\ee
where $Q$ here is summed over the box modes in a single direction under a periodic boundary condition. %
This choice is the most general ansatz considered in this work, and as we will show, the most energetically favourable FFLO phase.

For the sake of completeness, we mention that the multi-directional symmetry breaking phase transition is %
referred to as LO$_{\text{2}}$. When the system experiences a large population imbalance the order parameter %
may depend on two or three directions in space. It can be argued that for symmetry reasons, similar to the LO phase discussion, %
the LO$_{\text{2}}$ transition should occur with ${Q}_i=Q$ in any $i^{\text{th}}$ direction, leading to a form %
$\Delta_{\text{LO}_{\text{2}}}(\vect{x})=\Delta_{Q}(\cos(Qx)+\cos(Qy))$ in two and three dimensions, or %
$\Delta_{\text{LO}_{\text{3}}}(\vect{x})=\Delta_{Q}(\cos(Qx)+\cos(Qy)+\cos(Qz))$ in three dimensions. The role %
played by these phases involves only a small part of the phase diagram at large polarization~\cite{PhysRevA.87.063612}, %
and we can neglect them from the main study of the \LOg~phase.  

\section{\label{sec:model}Model}

In this work we consider a polarized Fermi gas in the 2D regime through a single channel model. Highly oblate systems %
can be realized via a combination of harmonic traps along the radial and axial directions, where %
 the anisotropic aspect ratio is $\lambda=\omega_z/\omega_\rho$, for axial confinement, $\omega_z$, and the radial counterpart %
denoted by $\omega_\rho$. %
When the temperature of the gas, $k_B T$, is significantly smaller than the energy spacing %
of the harmonic oscillator states, $\hbar\omega_z$ or $\hbar\omega_\rho$, the condition %
$k_BT\ll\hbar\omega_z\ll\hbar\omega_\rho/N$ describes a 1D regime with the transverse %
direction frozen, while  the condition
\begin{equation}\label{eq:2d_regime}
k_BT\ll\hbar\omega_\rho\ll\hbar\omega_z/\sqrt{N}
\end{equation}  
freezes out the axial direction and describes a 2D system. 
The  2D Fermi gas is then described by the following effective single channel Hamiltonian, 
\begin{equation}\label{eq:hamiltonian}
\begin{split}
\mathcal{H}=&\sum_\sigma\left[\int_V d^2x\ \psi^*_\sigma(\vect{x})\left(-\frac{\hbar^2}{2m}\vect{\nabla}^2 %
-\mu_\sigma\right)\psi_\sigma(\vect{x})\right]\\
&+g_{\text{2D}}\int_V d^2x\ \psi^*_\uparrow(\vect{x})\psi^*_\downarrow(\vect{x})\psi_\downarrow(\vect{x})\psi_\uparrow(\vect{x}),
\end{split}
\end{equation}
where $\psi_\sigma(\vect{x})$ is the Fermi field operator  with pseudospins $\sigma=\uparrow,\downarrow$, $V$ is the area of %
the system, $g_{\text{2D}}$ the magnitude of the contact interaction, $\mu_{\uparrow,\downarrow}=\mu\pm\delta\mu$ the chemical potentials for the spin components, %
$\delta\mu$ the chemical potential imbalance, and $m$ the mass of the atoms. 
We define the polarization of the system in terms of the populations $N_\uparrow$ and $N_\downarrow$,
\begin{equation}
p=\frac{N_\uparrow-N_\downarrow}{N_\uparrow+N_\downarrow},
\end{equation}  
where $N=N_\uparrow+N_\downarrow$ is the total number of atoms in the system. The interaction is modeled as a contact potential in the {\it s}-wave channel,
$U_{\text{2D}}(\mathbf{x},\mathbf{x}')=g_{\text{2D}}\delta(\vect{x}-\vect{x}')$ and in two dimensions must be carefully dealt with. The contact interaction should be renormalized due to the unphysical ultraviolet divergence (UV) at high energy. From the two-body $T$-matrix we have
\be
T^{-1}(E) = Vg^{-1}_{\text{2D}}-\Pi(E),
\ee 
where $E$ is the scattering energy in the center-of-mass frame, the particle-particle bubble is given by
\be
\Pi(E) = \sum_{\mathbf{k}}\frac{1}{E-2\epsilon_{\mathbf{k}}+i0^+},
\ee 
and $\epsilon_{\mathbf{k}} = \hbar^2\mathbf{k}^2/2m$.
The particle-particle bubble diverges logarithmically in two dimensions and we regularize this by introducing a cut-off, $\Lambda$, for large $k$.
We match the {\it s}-wave 2D scattering amplitude to the two-body $T$-matrix, 
\begin{equation}
T(E)=\frac{4\pi\hbar^2/m}{\log(\varepsilon_B/E)+i\pi},
\end{equation}
in the zero energy limit and for the binding energy $\varepsilon_B$ of the two-body bound state. 
The scattering amplitude can be recovered from the contact potential by setting and defining,  %
\begin{equation}
\frac{1}{g_{\text{2D}}(\Lambda)}=-\frac{1}{V}\sum_{|\vect{k}|<\Lambda}\frac{1}{2\epsilon_{\mathbf{k}}+\epsilon_B}.
\end{equation}
The two-body binding energy is then related to the 2D scattering length via
\begin{equation}
\epsilon_B=\frac{\hbar^2}{ma^2_{\text{2D}}}.
\end{equation}
It is useful to compare 2D scattering to its 3D counterpart, describing how a quasi-2D system is related to a 3D system \cite{Petrov2dscatt,Levinsen2015}. 
 Integrating out the  the %
axial direction we can relate the 3D scattering length to an effective quasi-2D scattering length, 
\begin{equation}\label{eq:2D3D}
\frac{a_{\text{2D}}}{l_z}=\sqrt{\frac{\pi}{B}}\exp\left(-\sqrt{\frac{\pi}{2}}\frac{l_z}{a_{\text{3D}}}\right),
\end{equation} 
where $l_z=\sqrt{\hbar/(m\omega_z)}$ is the axial harmonic oscillator length and %
$B\simeq0.905$ is a constant. Equation~\eqref{eq:2D3D} is valid when the scattering energy is %
neglegible compared with the strength of the confinement.
In actual experiments, the scattering length, $a_{\text{2D}}$, can be changed via Feshbach resonances and %
hence the effective interactions can be tuned~\cite{Dyke2016}. %

\subsection{\label{subsec:BdG}BdG self-consistent method}

We now turn to the many-body formulation of a population imbalanced Fermi gas considered through the BdG equations. %
Starting from the Heisenberg equation of motion of the Hamiltonian found in Eq.~\eqref{eq:hamiltonian}, we define %
$\sgn(\uparrow)=-\sgn(\downarrow)=1$ and we obtain for the fermionic fields $\psi_\sigma(\vect{x},t)$,
\be
\im\partial_t\psi_\sigma=\left(-\frac{\hbar^2\vect{\nabla}^2}{2m}-\mu_\sigma\right)\psi_\sigma+%
g_{2\text{D}}(\sgn\sigma)\psi^*_{\bar{\sigma}}\psi_\downarrow\psi_\uparrow.
\ee
Via the mean-field approximation, we can replace the coupled terms, $g_{\text{2D}}\psi^{\dagger}_{\uparrow}\psi_\downarrow\psi_\uparrow$ and $g_{\text{2D}}\psi^{\dagger}_{\uparrow}\psi_\downarrow\psi_\uparrow$ with their respective mean-fields,  
\be
g_{2\text{D}}(\sgn\sigma)\psi^*_{\bar{\sigma}}\psi_\downarrow\psi_\uparrow\simeq g_{2\text{D}}n_{\bar{\sigma}}(\vect{x})\psi_{\sigma}%
-(\sgn\sigma)\Delta(\vect{x})\psi^*_{\bar{\sigma}},
\ee
where we have defined the order parameter and the density profiles, 
$\Delta(\vect{x})=-g_{\text{2D}}\langle\psi_\downarrow(\vect{x})%
\psi_\uparrow(\vect{x})\rangle$ and $n_{\sigma}(\vect{x})=\langle\psi^{\dagger}_\sigma(\vect{x})%
\psi_\sigma(\vect{x})\rangle$. 
The above decoupling of the fields gives the following to the equations of motion for the spin component $\sigma$,
\be
i\frac{\partial\psi_\sigma}{\partial t}(\vect{x},t) = (\mathcal{H}_0 - \mu_\sigma)\psi_\sigma(\vect{x},t) - \sgn(\sigma)\Delta(\vect{x})\psi^{\dagger}_{\bar\sigma}(\vect{x},t),
\ee
where $\mathcal{H}_0=-\hbar^2\nabla^2_{\vect{x}}/(2m)$. Note that we will discard the Hartree term as we take $g_{\text{2D}}\rightarrow 0$ in the renormalization procedure. We insert the standard Bogoliubov transformation to solve the equations of motion,
\be\begin{split}
\psi_\uparrow(\vect{x},t)&=%
\sum_\eta\left[u_{\eta\uparrow}(\vect{x})c_{\eta\uparrow}e^{-\im E_{\eta\uparrow} t}-%
v^*_{\eta\downarrow}(\vect{x})c^{\dagger}_{\eta\downarrow}e^{\im E_{\eta\downarrow} t}\right], \nonumber \\
\psi^{\dagger}_\downarrow(\vect{x},t)&=%
\sum_\eta\left[u^*_{\eta\downarrow}(\vect{x})c^{\dagger}_{\eta\downarrow}e^{\im E_{\eta\downarrow} t}+%
v_{\eta\uparrow}(\vect{x})c_{\eta\uparrow}e^{-\im E_{\eta\uparrow} t}\right],
\end{split}\ee
where the wave functions $u_{\sigma\eta}(\vect{x})$ and $v_{\sigma\eta}(\vect{x})$ are normalized by the condition
\be
\int_V d\mathbf{x} \,(|u_{\sigma\eta}(\vect{x})|^2+|v_{\sigma\eta}(\vect{x})|^2)=1.
\ee
This gives the well known BdG equations,
\begin{alignat}{1}\label{eq:bdg}
\smash{\fourvec{\mathcal{H}_0-\mu_\sigma}{-\Delta(\vect{x})}{-\Delta^*(\vect{x})}{-\mathcal{H}_0+\mu_{\bar{\sigma}}}
\twovec{u_{\eta\sigma}(\vect{x})}{v_{\eta\sigma}(\vect{x})}= E_{\eta\sigma}
\twovec{u_{\eta\sigma}(\vect{x})}{v_{\eta\sigma}(\vect{x})} },
\end{alignat}
for quasiparticle energy $E_{\eta\sigma}$. 
The unequal spin populations require the chemical potential of each spin component to be unequal, this leads to different quasiparticle wave functions for each spin component, however, the $\downarrow$ problem is related to the $\uparrow$ problem through the unitary transformation %
\be
\twovec{u_{\eta\sigma}(\vect{x})}{v_{\eta\sigma}(\vect{x})}%
=
\twovec{-v^*_{\eta\bar{\sigma}}(\vect{x})}{u^*_{\eta\bar{\sigma}}(\vect{x})},
\ee
and the real-valued quasiparticle energies through, $E_{\eta\sigma}=-E_{\eta\bar{\sigma}}$. Without loss of generality we then remove the spin index of the BdG equations, defining %
$u_\eta=u_{\eta\uparrow}$, $v_\eta= v_{\eta\uparrow}$ and $E_\eta=E_{\eta\uparrow}$, leading to the following form 
\be\label{eq:eigenprob2}
\fourvec{\mathcal{H}_0-\mu_\uparrow}{-\Delta(\vect{x})}{-\Delta^*(\vect{x})}{-\mathcal{H}_0+\mu_{\downarrow}}
\twovec{u_{\eta}(\vect{x})}{v_{\eta}(\vect{x})}= E_\eta
\twovec{u_{\eta}(\vect{x})}{v_{\eta}(\vect{x})}.
\ee
Through the Bogoliubov transforms the density profiles %
$n_\sigma(\mathbf{x})$ can be written in terms of the quasi-particle creation and annihilation operators, 
\begin{gather}
\label{eq:nupdef}n_\uparrow(\vect{x}) = \sum_\eta u^{*}_\eta(\vect{x})u_\eta(\vect{x}) f(E_\eta), \\
\label{eq:ndowndef}n_\downarrow(\vect{x}) = \sum_\eta v^{*}_\eta(\vect{x})v_\eta(\vect{x}) f(-E_\eta),
\end{gather}
and the pairing gap function becomes, 
\be
\label{eq:profiledef}\Delta(\vect{x}) = -g_{\text{2D}}\sum_\eta u_{\eta}(\vect{x})v^*_\eta(\vect{x}) f(E_\eta).
\ee
The contact interaction introduces a UV divergence in the order parameter and this will be dealt with in the next section. We impose fermionic statistics for excitations  at inverse temperature $\beta=1/(k_BT)$ through the Fermi-Dirac distribution 
$f(E_{\eta})=1/(1+e^{\beta E_{\eta}})$.

\subsection{\label{subsec:BdG_LOg} BdG equations with the \LOg~ansatz}

The first step to solve the BdG equations is to expand the quasiparticle wave functions onto a complete basis and define the order parameter as being a real valued function. The complete basis we will use is a tensor product of a 1D complete orthonormal basis that satisfies the periodic boundary conditions of a box with length $L$. Specifically,  for a given $\vect{n}=(n_1,n_2)\in\NN\times\NN$,  we have
\be \label{Eq:basisprod}
\phi^{(2)}_{\vect{n}} (\vect{x})= \phi^{(1)}_{n_1}(x)\otimes\phi^{(1)}_{n_2}(y),
\ee
where,
\be
\phi^{(1)}_n(x)=\left\{
\begin{array}{lll}
L^{-1/2} & & \text{if }n=0, \\
(2/L)^{1/2}\cos\left[n\pi x/L\right] & & \text {if $n$ is even,} \\
(2/L)^{1/2}\sin\left[(n+1)\pi x/L\right] & & \text {if $n$ is odd,}
\end{array}
\right.
\ee 
and $n_1$ and $n_2$ are non-negative integers. Letting $\hbar=2m=1$, %
the 1D basis functions, $\phi^{(1)}_n(x)$, are eigenvectors %
for the one-dimensional free Hamiltonian, $\mathcal{H}=-(d^2/dx^2)$, with eigenvalues 
\be
\epsilon_n\defeq\left\{
\begin{array}{lll}
\pi^2n^2/L^2 & & \text {if $n$ is even}, \\
\pi^2(n+1)^2/L^2 & & \text {if $n$ is odd}.
\end{array}
\right.
\ee
The product basis, Eq.~\eqref{Eq:basisprod}, are eigenvectors for $\mathcal{H}_0$ with eigenvalues $\epsilon_{\vect{n}}=\epsilon_{n_1}+\epsilon_{n_2}$.  
We expand the basis functions $u_\eta(\mathbf{x})$ and $v_\eta(\mathbf{x})$ onto this basis
$$
u_\eta(\vect{x}) = \sum_{\vect{n}} A^{(\eta)}_{\vect{n}}\phi^{(2)}_{\vect{n}}(\vect{x}),\qquad %
v_\eta(\vect{x}) = \sum_{\vect{m}} B^{(\eta)}_{\vect{m}}\phi^{(2)}_{\vect{m}}(\vect{x}), 
$$
and the BdG equations, Eq.~\eqref{eq:eigenprob2}, becomes %
\begin{alignat}{1}\label{eq:eigenprob}
\fourvec{(\varepsilon_{\mathbf{n}}-\mu_\uparrow)\delta_{\mathbf{nm}}}{-\Delta_{\mathbf{nm}}}{-\Delta^*_{\mathbf{nm}}}{-(\varepsilon_{\mathbf{n}}-\mu_\downarrow)\delta_{\mathbf{nm}}}
\twovec{A^{(\eta)}_{\vect{m}}}{B^{(\eta)}_{\vect{m}}}
=E_{\eta}
\twovec{A^{(\eta)}_{\vect{n}}}{B^{(\eta)}_{\vect{n}}},
\end{alignat}
where, 
\begin{equation}\label{eq:manddelta}
\Delta_{\vect{n}\vect{m}} = \int_Vd^2x\ \phi^{(2)}_{\vect{n}}(\vect{x})\Delta(\vect{x})\phi^{(2)}_{\vect{m}}(\vect{x}).
\end{equation}

The eigenvalue problem described by Eq.~\eqref{eq:eigenprob} comes at significant computational cost and is not easily tractable. In this work we solve the simpler case of an order parameter %
$\Delta(\vect{x})$ that is a real-valued function depending on the $x$ direction only and is a sum of $Q$ modes, the 
\LOg~ansatz defined in Eq.~\eqref{eq:lo_g_ansatz}. The periodic boundary conditions force the Fourier transform of %
$\Delta(\vect{x})$ to be dependent on %
the discretized, even modes of the box. This fixes $Q=2\pi n/L$ for each non-negative integer $n\in\NN$. %
From Eq.~\eqref{eq:manddelta}, 
we integrate for each $\vect{n}=(n_1,n_2)$ and each $\vect{m}=(m_1,m_2)$ over the variable $y$, i.e., 
\bea
\Delta_{\vect{n}\vect{m}} &=& \delta_{n_2,m_2}\int_{-L/2}^{L/2}dx\ %
\phi^{(1)}_{n_1}(x)\Delta(x)\phi^{(1)}_{m_1}(x), \nonumber\\
&=&\delta_{n_2,m_2}\Delta_{n_1,m_1}.
\eea
We can simplify the BdG equations further by exploiting the $y$ dependence of the densities and order parameter for all $\eta$ in Eqs.~\eqref{eq:nupdef}, \eqref{eq:ndowndef}, \eqref{eq:profiledef}, and the BdG equations, Eq.~\eqref{eq:eigenprob}. We see that the dependence on $m_2$ falls out of the problem and the BdG equations become block diagonal in $n_2$. This simplifies the BdG equations and we have for our final form,
\begin{alignat}{1} \label{Eq:bdgfinal}
\fourvec{\varepsilon_{\uparrow} }{-\Delta_{n_1m_1}}{-\Delta_{n_1m_1}}{-\varepsilon_{\downarrow}}
\twovec{A^{(\eta n_2)}_{m_1} }{B^{(\eta n_2)}_{m_1}} 
=E_{\eta}^{n_2}
\twovec{A^{(\eta n_2)}_{n_1} }{B^{(\eta n_2)}_{n_1}}.
\end{alignat}
where $\varepsilon_{\uparrow} = (\varepsilon_{n_1m_1,n_2}-\mu_\uparrow)\delta_{n_1m_1}$ and $\varepsilon_{\downarrow} = (\varepsilon_{n_1m_1,n_2}-\mu_\downarrow)\delta_{n_1m_1}$.

\subsection{\label{subsec:implementation} Hybrid BdG strategy}

In order to deal with the infinite dimensional size of the BdG equations matrix we introduce a cut-off energy $E_c$ to make the matrix equations finite, and treat %
the high-lying energy states in Eqs.~\eqref{Eq:bdgfinal} separately 
as free states, or plane-wave functions~\cite{Liu20071D}. Specifically, we have the discrete part, below %
the cut-off, and a continuous part above the cut-off. From Eqs.~\eqref{eq:nupdef},~\eqref{eq:ndowndef}, %
and~\eqref{eq:profiledef} we truncate the sums to obtain, 
\begin{gather}
n_{d\uparrow}(\vect{x}) = \sum_{|E_\eta|<E_c} u^*_\eta(\vect{x})u_\eta(\vect{x}) f(E_\eta), \\
n_{d\downarrow}(\vect{x}) = \sum_{|E_\eta|<E_c} v^*_\eta(\vect{x})v_\eta(\vect{x}) f(-E_\eta), \\
\Delta_d(\vect{x}) = -g_{\text{2D}}\!\!\!\!\sum_{|E_\eta|<E_c} u_{\eta}(\vect{x})v^*_\eta(\vect{x})f(E_\eta), 
\end{gather}
and where the gap equation is still yet to be renormalized. The high-lying states, treated as plane waves, 
\begin{alignat}{1}
u_\eta(\vect{x}) \rightarrow \frac{1}{\sqrt{V}}u_k(\vect{x}) e^{\im \vect{k}\cdot\vect{x}}, \nonumber \\
v_\eta(\vect{x}) \rightarrow \frac{1}{\sqrt{V}}v_k(\vect{x}) e^{\im \vect{k}\cdot\vect{x}},
\end{alignat}
allows us to solve \eqref{Eq:bdgfinal} and compute the energies above the cut-off. In particular, we %
can define the energies $E_k(\vect{x})=\sqrt{(\epsilon_k-\mu)^2+\Delta(\vect{x})^2}$ and obtain %
the continuum corrections,
\begin{alignat}{1}
n_{c\uparrow}(\vect{x}) &= \sum_{\vect{k}}%
\left(\frac{1}{2}-\frac{\epsilon_k-\mu}{2E_k(\vect{x})}\right)%
\Theta(E_k(\vect{x})+\delta\mu - E_c), \\
n_{c\downarrow}(\vect{x}) &= \sum_{\vect{k}}%
\left(\frac{1}{2}-\frac{\epsilon_k-\mu}{2E_k(\vect{x})}\right)%
\Theta(E_k(\vect{x})-\delta\mu - E_c), \\
\Delta_c(\vect{x}) &= -g_{\text{2D}}\sum_{\vect{k}}\frac{\Delta(\vect{x}) }{2E_k(\vect{x})}\Theta(E_k(\vect{x})+\delta\mu - E_c),
\end{alignat}
where $\Theta(x)$ is the Heaviside step function. The continuous and discrete parts of the order parameter can be combined and using the regularization condition we have, 
\begin{alignat}{1}
\Delta(\vect{x}) =-g_{\text{2D}}^{\text{eff}}(\vect{x})\sum_{|E_\eta|<E_c} u_{\eta}(\vect{x})v^*_\eta(\vect{x})f(E_\eta),
\end{alignat}
where,
\be
\frac{1}{g_{\text{2D}}^{\text{eff}}(\vect{x})}=\frac{1}{g_{\text{2D}}}+g(\vect{x}),
\ee
and using the definition, 
\be
g(\vect{x})= \frac{1}{V}\sum_{\vect{k}}\frac{1}{2E_k(\vect{x})}\Theta(E_k(\vect{x})+\delta\mu - E_c).
\ee
In the thermodynamic limit, due to the high density %
of states beyond the cut-off, the continuous contribution can be analytically solved, giving,
\begin{gather}\label{eq:highen}
n_{c\uparrow}(\vect{x}) = \frac{1}{8\pi}\left[E_c-\delta\mu +\mu - k_c(\vect{x})^2\right], \\
n_{c\downarrow}(\vect{x}) = \frac{1}{8\pi}\left[E_c+\delta\mu +\mu - k_c(\vect{x})^2\right], \\
\frac{1}{g_{\text{2D}}^{\text{eff}}(\vect{x})} = %
-\frac{1}{8\pi}\log\left(\frac{E_c-\delta\mu-\mu+k_c(\vect{x})^2}{\epsilon_B}\right),
\end{gather}
where we introduce the cut-off momentum, $k_c(\vect{x})=\left(\mu+\sqrt{(E_c-\delta\mu)^2-\Delta(\vect{x})^2}\right)^{1/2}$.
Gathering together all these results we achieve the final form for the densities profiles, 
\be\begin{split}
n_\uparrow (\vect{x})&= n_{d\uparrow}(\vect{x}) + n_{c\uparrow}(\vect{x}),\\
n_\downarrow (\vect{x})&= n_{d\downarrow}(\vect{x}) + n_{c\downarrow}(\vect{x}),\\
\Delta(\vect{x})&=-g_{\text{2D}}^{\text{eff}}(\vect{x})\Delta_d(\vect{x}).
\end{split}\ee

In the grand canonical ensemble we find the absolute minimum of the thermodynamic potential, %
$\Omega=\Omega(\mu,\delta\mu,T)=U-TS-\mu N-\delta\mu\delta N$, where $U$ is the internal energy and $S$ is the entropy. %
Simulations performed in the canonical ensemble requires more computational effort, where the BdG equations depend self-consistently on $\mu$ and $\delta\mu$ that are computed using the number equations 
\be\label{eq:number_eq}
N=-\frac{\partial\Omega}{\partial\mu}, \qquad \delta N=-\frac{\partial\Omega}{\partial(\delta\mu)},
\ee
converging to a given $\mu(n,\delta n, T)$ and $\delta\mu(n,\delta n, T)$.  To find the ground state for momenta $Q$ we minimize the Helmholtz free energy, %
$F=F(n,p,T)=U-TS$ with respect to $Q$.

\subsection{Energy and Entropy}

In addition to the calculation of the order parameter and densities, the complete study of this system, either at zero or finite temperature, requires the total energy and entropy densities.  
Through the Bogoliubov transformations we define the profiles for the energy density, $E(\vect{x})$, %
and the entropy density, $S(\vect{x})$, on quasiparticles modes as 
\begin{alignat}{1}
\label{eq:endef}
E(\vect{x}) = & \mu_\uparrow n_\uparrow(\vect{x})+\mu_\downarrow n_\downarrow(\vect{x})-\frac{|\Delta(\vect{x})|^2}{g}  \nonumber \\  
+ & \smash{\sum_\eta\left[(|u_\eta(\vect{x})|^2+|v_\eta(\vect{x})|^2)f(E_\eta)-|v_\eta(\vect{x})|^2\right]},  \\
\label{eq:entropydef}
S(\vect{x}) = & -k_B\sum_\eta\bigl[f(E_\eta)\log(f(E_\eta))  \nonumber \\  +&  f(-E_\eta)\log(f(-E_\eta))\bigl]%
(|u_\eta(\vect{x})|^2+|v_\eta(\vect{x})|^2).
\end{alignat}
Since both the energy density and entropy density %
depend on sums over the eigenvalues, $E_\eta$, and eigenvectors, $u_\eta(\vect{x})$ and $v_\eta(\vect{x})$, we can calculate the energy and entropy densities of the system through the expansion of the discrete and continuous states. 
It is possible to show that the continuous correction to the entropy density is negligible for high-lying states \cite{Liu2007}, and we focus on the energy.
Treating the high energy states %
in the thermodynamic limit, we write $E(\vect{x})=E_d(\vect{x})+E_c(\vect{x})$, and get the following contribution, 
\begin{widetext}
\begin{alignat}{1}
E_c(\vect{x})
=\frac{1}{8\pi}\left\{\Delta^2\left[\frac{1}{2}+\log\left(\frac{\sqrt{E_-+\sqrt{E_-^2-\Delta^2}}%
\sqrt{E_++\sqrt{E_+^2-\Delta^2}}}{\epsilon_B}\right)\right]%
+\sum_{\sigma=\pm}\frac{E_\sigma}{2}\sqrt{E_{\bar{\sigma}}^2-\Delta^2}-E_c^2+\delta\mu^2\right\},
\end{alignat}
\end{widetext}
where for simplicity we define $E_\pm=E_c\pm \delta\mu$. 

The solutions to the BdG equations were carried out following an iterative algorithm, %
self-consistently calculating the order parameter and the density profiles. We perform calculations in either %
the grand canonical ensemble, fixing $\mu$ and $\delta\mu$ and computing the number %
of particles $N$ and polarization $p$, or in the canonical ensemble, fixing $N$ and $p$, and %
iteratively computing the $\mu$ and $\delta\mu$. 
It is important to remark that the nature of the quantities we want to %
investigate, the Helmholtz free energy and thermodynamic potential, show very small variations while %
changing the thermodynamic variables. From the knowledge we have from solving the FF ansatz found in %
Fig.~\ref{fig:tp_ff}, the relative difference needed to see the appearance of new absolute minima in the %
thermodynamic potential is approximately $10^{-7}$. To reach this precision we have to increase both the %
number of particles, which increases the size of BdG equations, and the sampling rate over the box length, %
which increases the integration precision. 
Once the algorithm reaches convergence, the cut-off energy dependence is tested ensuring the %
calculations do not depend on the cut-off $E_c$.

The BdG equations are a mean-field treatment of the imbalanced system and they are expected to be quantitatively incorrect in the strongly correlated regime. However, they will provide constant qualitative results for all interaction strengths, and will provide an excellent picture of the system under investigation.

\section{\label{sec:results}Results}

In the calculations we use natural %
units, such that $\hbar=2m=k_B=1$ and, in the case of a non-vanishing polarization, we set $k_F=1$ where $k_F$ is, at %
fixed particle number, the free Fermi momentum. In 2D the linear relation between the Fermi energy and the density of particles requires that %
the size of the box $L$ increases as the square root of the number of particles, $N$. An increase in the number of %
particles corresponds to a quadratic increase in the size of the BdG equations. We have used a particle number of %
$N=2000$, which corresponds to a box size of $L\simeq 112/k_F$, comparable %
with similar results in the 1D case~\cite{Liu2007}. 
We set the cut-off energy for values $E_c\geq 10\epsilon_F$, where $\epsilon_F=\hbar^2k_F^2/2m$ %
is the Fermi energy, however some calculations require higher cut-off values. It is interesting to note that, for the self-consistent calculations, a large cut-off, $E_c$, slows down the runtime, but reduces the number of %
iterations before convergence. %
The accuracy of the diagonalization process of the BdG equations %
has been implemented with an additional number, $\bar{n}$, of orthonormal basis elements beyond the cut-off energy, %
such that
\be\label{eq:sqrt}
n_1<\sqrt{\frac{E_c L^2}{\pi^2}-n^2_2}=\bar{n}.
\ee

In the following subsections we will look at the results for the \LOg~phase in the canonical ensemble and follow with the treatment of the grand canonical ensemble results. We will discuss the phase diagram at zero and finite temperature, the structure of %
the \LOg~ground states that depends on the spin imbalance. A specific comparison with %
the LO$_{1}$ ansatz will be treated. The behaviour in the grand canonical ensemble will be extensively investigated %
with emphasis on the order of the phase transition.  

\subsection{Canonical ensemble}
\begin{figure}
\includegraphics[width=0.465\textwidth]{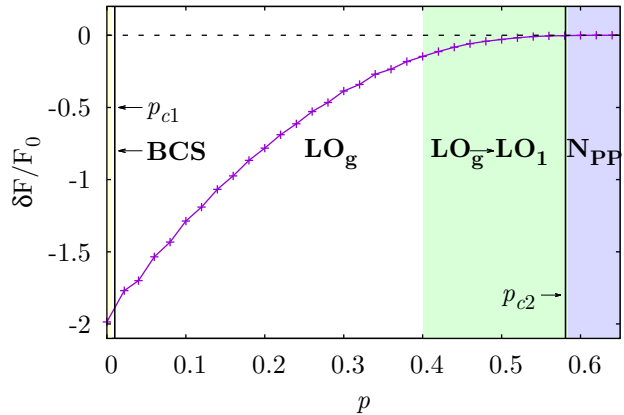}
\caption{\label{fig:dm_p}(color online). The energy difference $\delta F=F_{\LOgf}-F_{\text{free}}$, for the  Helmholtz free energy, $F_{\LOgf}$, %
and the free Fermi gas, $F_{\text{free}}$, at temperature $T/T_F=0$, and interaction strength $\varepsilon_B=0.25\varepsilon_F$. Re-scaled for graphic %
purposes through the constant $F_0=2\pi\times10^{-2}N\epsilon_F$. The transition polarization %
$p_{c1}\simeq 0.02$, from BCS to \LOg, and $p_{c2}=0.58$, from \LOg~to N$_{\text{PP}}$, %
are denoted by bold black vertical lines.
The region $p>0.4$ denotes %
the \LOg~phase that is indistinguishable from the LO$_1$ phase.}
\end{figure}

In Fig.~\ref{fig:dm_p} we present the \LOg~phase diagram in the canonical ensemble for polarizations, $p$, and for reduced temperature, $T/T_F=0$, %
and binding energy $\epsilon_B=0.25\epsilon_F$. For a fixed polarization $p$ we define the free Fermi gas Helmholtz free energy %
at zero temperature $F_{\text{free}}(T=0,n,p)$ and we introduce a rescaling constant %
$F_0=2\pi\times10^{-2}\epsilon_FN$, with $N$ the number of particles, %
such that we can plot the dimensionless quantity,  
\be
\delta F/F_0=(F_{\LOgf}(T,n,p)-F_{\text{free}}(T=0,n,p))/F_0.
\ee
We see that the \LOg~phase is the energetically favorable ground state (with respect to the BCS and FF minima) for polarizations %
$0.02\leq p\leq 0.58$, above $p>0.58$ the system falls into the free Fermi gas phase \footnote{It is possible the ground state for these polarizations is the more complicated LO$_2$ phase, which is not considered in this work.}. 
We know that a finite non-zero polarization destroys the BCS phase, and we see a transition %
at very low polarizations, $p_{c1}=0.02$. The finite nature of this transition is due to the numerical difficulty in finding the true ground state. We expect that for finite non-vanishing polarizations below the transition, $p<p_{c1}$, %
the \LOg~phase is still the ground state. At high polarizations $0.40<p<0.58$ the %
\LOg~and LO$_1$ phase have merged, and we will explore and elucidate this part of the phase diagram further below through the structure of the order parameter. 

\begin{figure}
\includegraphics[width=0.465\textwidth]{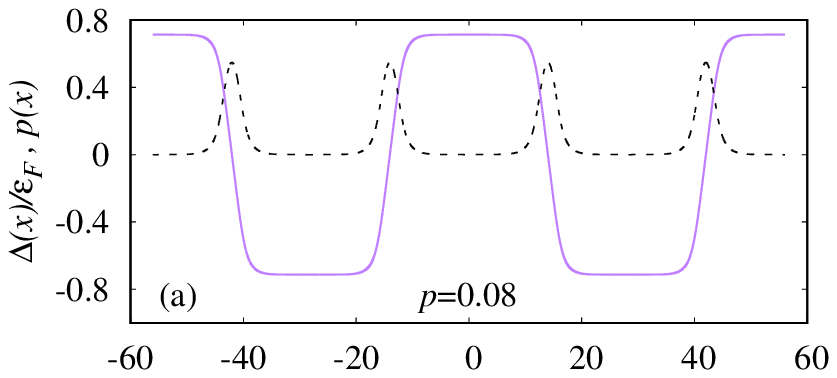}
\includegraphics[width=0.465\textwidth]{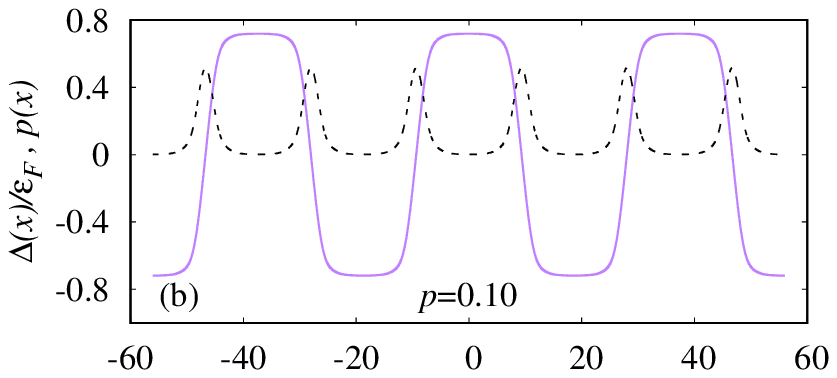}
\includegraphics[width=0.465\textwidth]{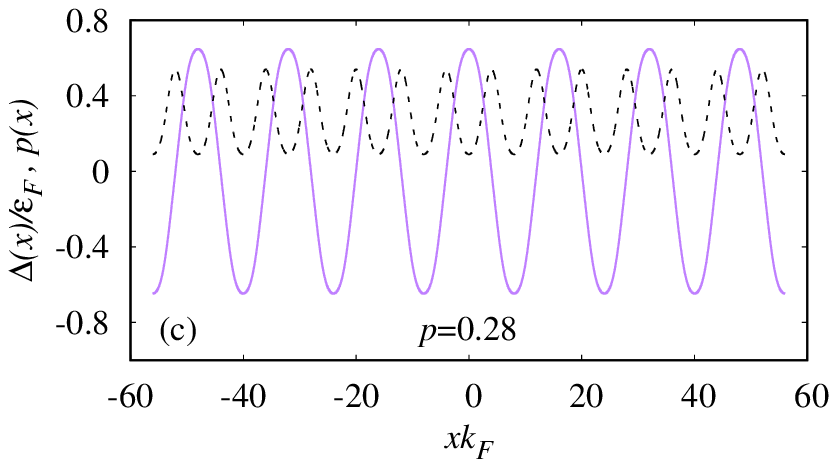}
\caption{\label{fig:delta_pola}(color online) The inhomogeneous order parameter $\Delta(x)/\epsilon_F$  %
as a function of the $x$ direction (solid line) %
and the local spin polarization $p(x)$ (dashed line) calculated via the self-consistent %
method at temperature, $T/T_F=0$, and interaction strength $\varepsilon_B=0.25\varepsilon_F$, for polarizations $p=0.08$ (a), $p=0.10$ (b), and $p=0.28$ (c).}
\end{figure}
In Fig.~\ref{fig:delta_pola} we explore the spatial distribution of the order parameter (solid line) and the local spin polarization (dashed line),
\be
p(x)=\frac{n_\uparrow(x)-n_\downarrow(x)}{n_\uparrow(x)+n_\downarrow(x)},
\ee
from the converged BdG solutions for polarizations $p=0.08$ (a), $p=0.10$ (b), and $p=0.28$ (c)  
as a function of the box length, $xk_F$, where $k_F$ is the Fermi momentum. We see in Figs.~\ref{fig:delta_pola}(a)-(c) that %
the increasing polarization, %
$p$, leads to a higher main amplitude, $Q$, and becomes consistent with the LO$_1$ ansatz. The local spin %
polarization is plotted as a function of space and we observe the structure imposed by the \LOg~phase. %
We see the favorable ground state configuration places the excess spin up atoms (the majority) where the order parameter vanishes, and the superfluid fraction at the peaks, and here excitations are less probable to occur, preserving superfluidity.

We can compare the \LOg~ground state order parameter with the LO$_{1}$ ansatz, by
decomposing the self-consistently computed order parameter through a cosine Fourier transform. The %
order parameter has to be an even function along the chosen preferred direction, that is 
\be\label{eq:ft}
\Delta_{\LOgf}(x)=\sum_{n=1}^{\infty}\Delta_n \cos\left(\frac{2\pi n}{L}x\right), 
\ee
then the \LOg~and the LO$_1$ are the same phase when the Fourier transform %
has only one mode. 
\begin{figure}
\centering
\includegraphics[width=0.465\textwidth]{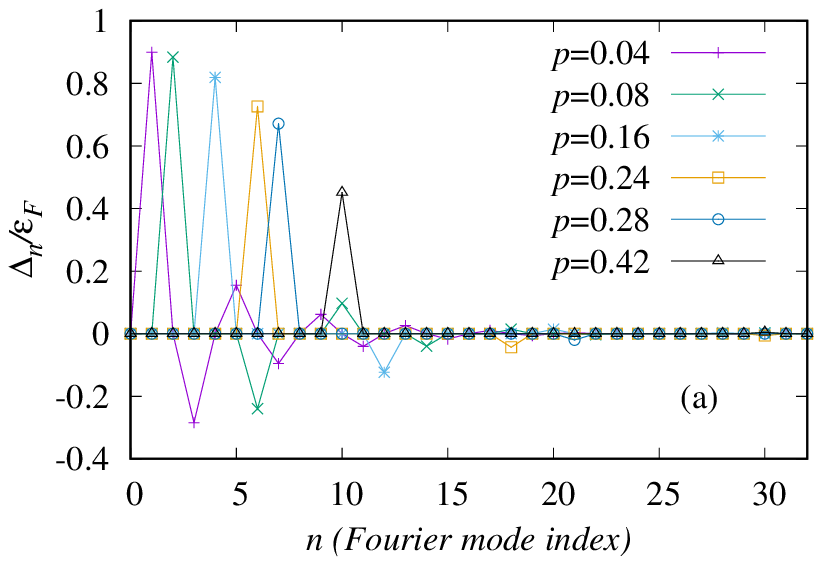}
\includegraphics[width=0.465\textwidth]{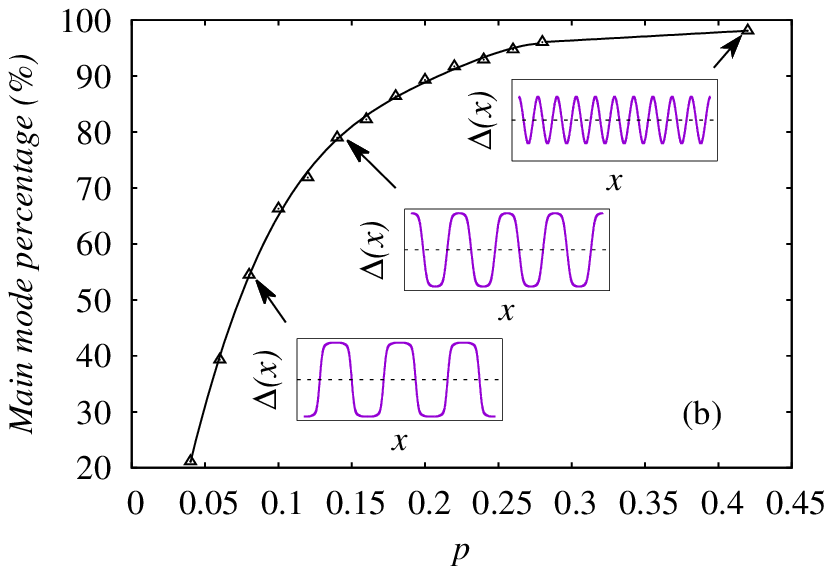}
\caption{\label{fig:combine}(color online) (a) The weight $\Delta_n$ (in Fermi energy units) of each Fourier cosine mode $n$ of the \LOg~phase, calculated according to Eq.~\eqref{eq:ft} at zero temperature and $\epsilon_B=0.25\epsilon_F$ , at different spin polarizations $p$. (b) The weight of the main Fourier mode, shown in percentage with respect to the other non-vanishing modes, as a function of the spin polarization. The insets are a guide to show the behavior of the order parameter for a given polarization. 
}
\end{figure} 
In Fig.~\ref{fig:combine}(a) we show that the self-consistent \LOg~phase is a superposition %
of different Fourier cosine modes and has a main mode occupation that is provided by the LO$_1$ ansatz. A comparison between the \LOg~phase configuration versus the LO$_1$ indicates that the \LOg~choice is always energetically favourable. This fact is %
true at both low and high polarizations where the \LOg~and LO$_1$ phases almost match. 
In order to %
explore this overlap we study the percentage occupation of the main mode %
and we observe in Fig.~\ref{fig:combine}(b) that a perfect overlap between the phases never fully occurs. %
We can then conclude that there is always a combination of modes that is the energetically preferred ground state.

\begin{figure}
\centering
\includegraphics[width=0.465\textwidth]{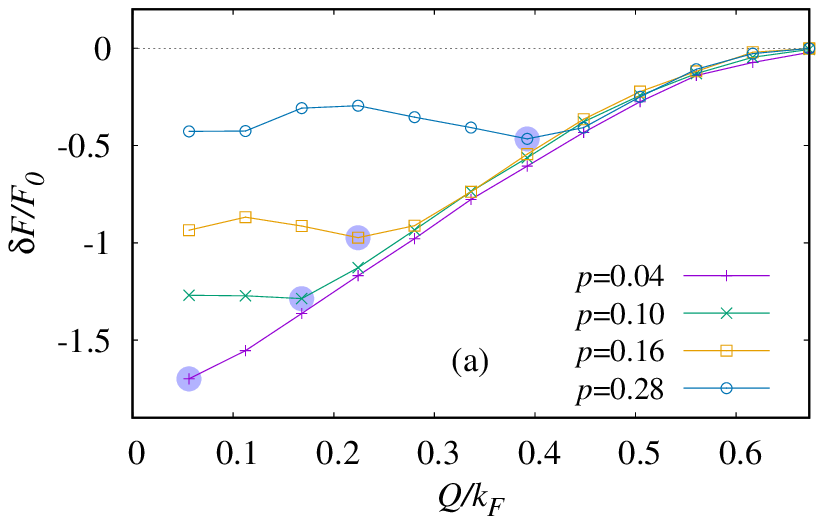}
\includegraphics[width=0.465\textwidth]{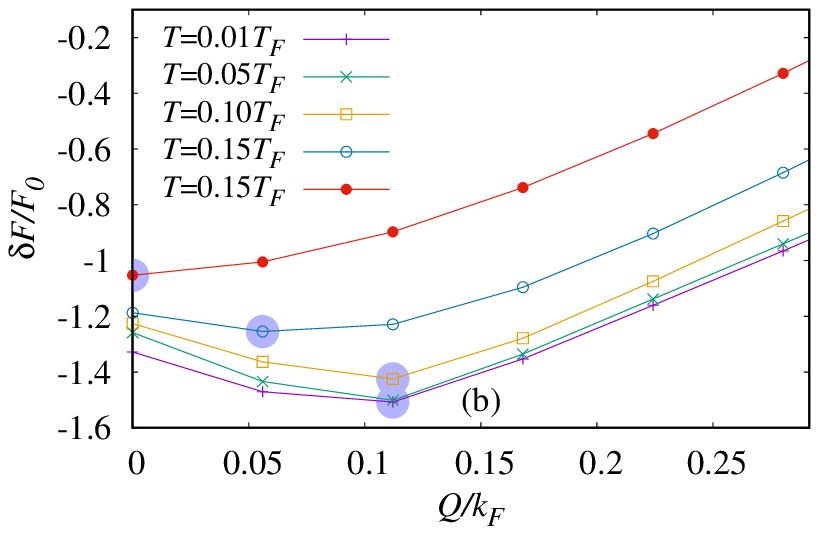}
\caption{\label{fig:helmholtz}(color online). The Helmholtz free energy difference between the \LOg~(for $Q\neq 0$ and BCS at $Q=0$) and %
the free Fermi gas free energy rescaled by $F_0=2\pi\times10^{-2}N\epsilon_F$: %
(a) at zero temperature varying the polarization and (b) at %
fixed polarization $p=0.07$ increasing the temperature. The semi-transparent bullets indicate %
the absolute minimum position $Q_\text{min}$ for each configuration. The dashed horizontal line in panel (a) %
is the free Fermi gas free energy: every time the $Q$ is too high the model mimics the %
free Fermi gas destroying the order parameter. Here, the binding energy is $\epsilon_B=0.25\epsilon_F$.}%
\end{figure} 

In Fig. \ref{fig:helmholtz} we show the behaviour of the Helmholtz free energy for two different cases, %
(a) at zero temperature and letting the polarization vary,  while in (b) the polarization is fixed at $p=0.07$ and we increase the temperature. 
Figure~\ref{fig:helmholtz}(a) shows that the minimum %
of the free energy that occurs at $Q_\text{min}$ always lies away from $Q/k_F=0$ and $Q_\text{min}$  increases proportionally %
with the polarization. Figure.~\ref{fig:helmholtz}(b) shows that there is a transition point between the \LOg~phase and the BCS phase in the temperature range %
$0.15T_F<T<0.20T_F$. We observe that the increase of polarization pushes the minimum to high %
$Q$ values, while the increase of temperature pushes it back towards the $Q=0$ limit (i.e., the BCS phase). This behaviour %
is expected to change for high polarization when the \LOg~phase has a transition directly to %
the free Fermi gas phase. We note that the interaction strength $\epsilon_B=0.25\epsilon_F$ requires a pair-fluctuation treatment to produce quantitative results as the mean-field theory predicts the transition temperature from the BCS to normal state of $0.41T_F$, well above the %
experimental observations \cite{Murthy2015,Boettcher2016}.


\subsection{Grand canonical ensemble}

Although the canonical ensemble can tell us about the properties and exotic phases of a spin polarized Fermi gas, %
it is important to know the nature of the phase transitions %
and the transition temperatures of the phases we have described so far; to perform this we will now work in the grand canonical ensemble. The absolute minimum of the thermodynamic potential is found at fixed chemical potentials, $\mu$ and $\delta\mu$, and this minimum moves through the landscape of the %
thermodynamic potential, giving us an insight as to the order of each phase transition. To explore the system we first need to study 
the behaviour of the the chemical potentials, $\mu$ and $\delta\mu$,  found through the self-consistent BdG method  in the canonical ensemble. We explore different regimes through the equations of state, $\mu(n,\delta n)$ and $\delta\mu(n,\delta n)$, with respect to a fixed density $n=k_F^2/(2\pi)$ and varying %
polarization to justify the regimes of interest in the grand canonical ensemble. 

\begin{figure}
\includegraphics[width=0.465\textwidth]{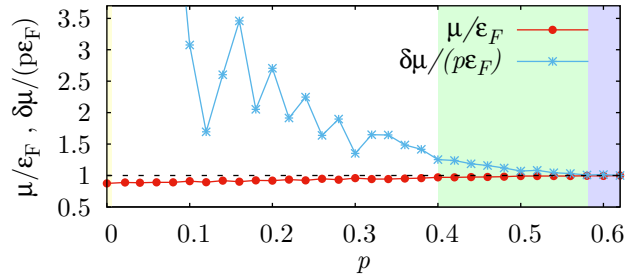}
\caption{\label{fig:m_p}(color online). The behaviour of $\mu$ and $\delta\mu$, with respect to their expected behaviour %
as free Fermi gas thermodynamic variables, as a function of the polarization, at zero temperature, density $n=k_F^2/(2\pi)$ and binding energy %
$\epsilon_B=0.25\epsilon_F$. The phase transition in the %
canonical ensemble are denoted by the background color as found in Fig.~\ref{fig:dm_p}.}
\end{figure}

The values of $\mu$ and $\delta\mu$ used %
to compute the real order parameter, $\Delta(x)$, at fixed density $n$ and %
polarization $p$, normalized by Fermi energy and polarization multiplied by the Fermi energy respectively, %
are plotted in Fig.~\ref{fig:m_p}. 
The shaded regions correspond to the phases found in Fig.~\ref{fig:dm_p}, at polarization $p_{c1}\simeq 0.02$, where we have the transition from BCS to \LOg, and $p_{c2}=0.58$, for the transition from \LOg~to N$_{\text{PP}}$ \footnote{The distinction between the fully polarized normal Fermi %
gas N$_{FP}$ and the partially polarized N$_{\text{PP}}$ that occurs when $\delta\mu>\mu$ is not taken into %
account here and we will use the notion of free Fermi gas for the N$_{\text{PP}}$ phase. }. 
At zero temperature, the BCS phase is favourable only at $p=0$. The underlying mean-field theory in 2D for the equation of state of a possible phase-separation phase (i.e., a mixture of a BCS superfluid and a normal gas) has been studied in Ref.~\cite{Lianyi2008} and %
the $\delta\mu$ versus $\mu$ phase diagram that involves the transition between BCS phase and the free %
Fermi gas has been explored therein. We take the salient results here,%
\be
\mu_{\text{BCS}}=\frac{\pi\hbar^2}{m}n-\frac{\epsilon_B}{2},\qquad 0\leq\delta\mu <\frac{\Delta_0}{\sqrt{2}}.
\ee
The upper limit for $\delta\mu$ is the Clogston-Chandrasekhar (CC) limit,
\be\label{eq:CC}
\delta\mu_{CC}=\sqrt{\frac{\epsilon_B^2}{2}+\mu\epsilon_B},
\ee
above which the superfluid breaks down.
We observe that %
both $\mu$ and $\delta\mu$ assume an almost direct proportionality with the Fermi energy $\epsilon_F$ and the %
polarization $p$ whenever $\delta\mu$ is large enough. Since in the grand canonical ensemble the \LOg~phase arises at high $\delta\mu$, %
we consider the case of $\mu/\epsilon_F=1$ and vary $\delta\mu$ through the CC limit at different %
binding energies. 

\begin{figure}
\includegraphics[width=0.465\textwidth]{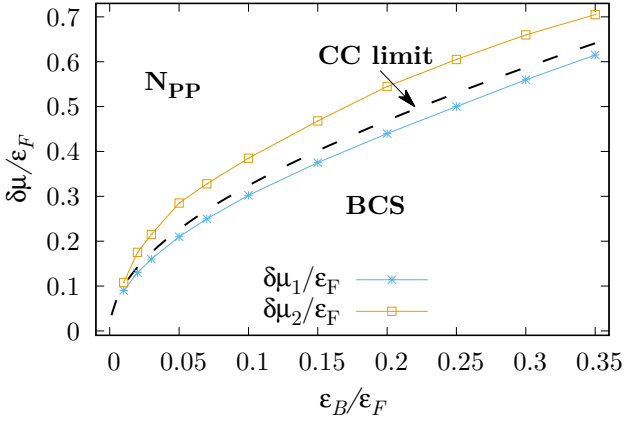}
\caption{\label{fig:dm_eb}(color online). The critical phase transition  values of $\delta\mu$ as a function of the %
binding energy, $\epsilon_B$, at zero temperature, %
$\delta\mu_1$ (stars) from the BCS superfluid to \LOg~and $\delta\mu_2$ (crosses)%
from \LOg~to \NPP.}
\end{figure}

Taking the above values for the chemical potentials, in Fig.~\ref{fig:dm_eb} we plot the phase diagram of the spin imbalanced Fermi gas, as a function of the binding energy, $\varepsilon_B/\varepsilon_F$, and dimensionalized chemical potential difference, $\delta\mu/\varepsilon_F$. The dashed line represents the CC limit, and we observe that the \LOg~phase provides %
two transition lines that we will denote with %
$\delta\mu_1$ (stars), below the CC limit, and $\delta\mu_2$ (squares), above it. In particular, the $\delta\mu_2$ line is in agreement with the results provided by the FF ansatz in Ref.~\cite{Sheehy2015}. It is interesting to observe how the reduced dimensionality of the problem, at least at zero temperature, %
enhances the effect of both the FF and the \LOg~phase. In 3D, it is possible to estimate that %
the FFLO phase is available for roughly 4\% of the $\delta\mu$ values %
for which a superfluid phase is favourable, either BCS or FF, as evaluated %
from the phase diagram temperature versus magnetization \cite{Parish2007}. We find from examining the corresponding region in  Fig.~\ref{fig:dm_eb} that,
at a fixed binding energy, the FFLO is accessible for roughly 17\% of the phase space, in which the system is a superfluid, %
that is, $(\delta\mu_2 -\delta\mu_1)/\delta\mu_2 \simeq 0.17$.

\begin{figure}
\includegraphics[width=0.415\textwidth]{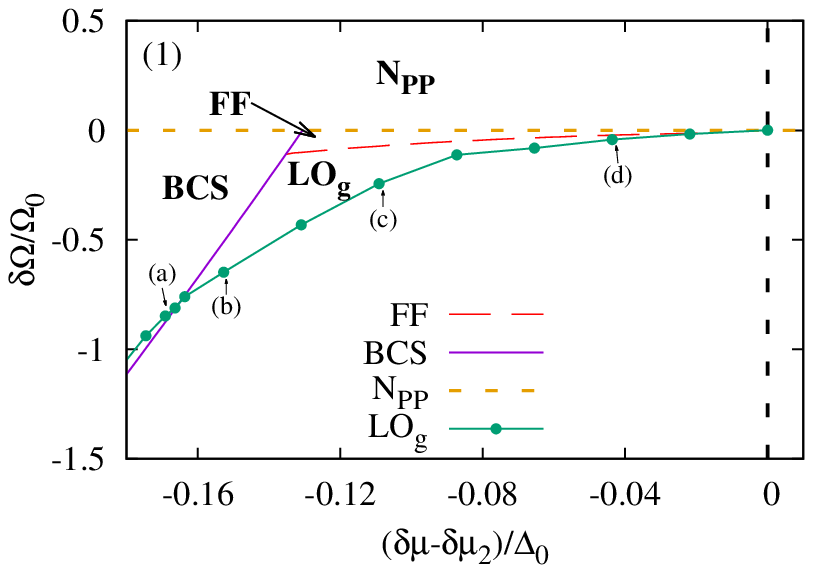}
\includegraphics[width=0.415\textwidth]{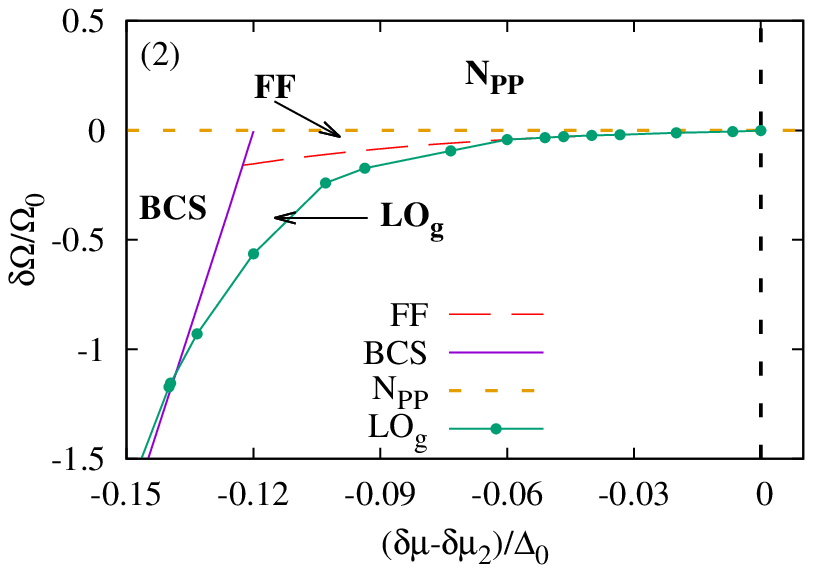}
\caption{\label{fig:tp_dm}(color online). The zero temperature minima of FF, BCS and \LOg~thermodynamic potentials. The upper panel (1) is for interaction strength $\epsilon_B=0.10\epsilon_F$ and panel (2) for interaction $\epsilon_B=0.25\epsilon_F$. The critical chemical potential imbalance, $\delta\mu_2$ denotes the \LOg~to \NPP~transition %
and $\Delta_0$ is the BCS energy gap at zero temperature. 
The thermodynamic potentials are defined with respect to the free Fermi gas thermodynamic potential, $\Omega_{\rm free}$, and scaled by the constant $\Omega_0 = 2\pi \times 10^{-3}N\varepsilon_F$.
In panel (1), the labels from (a) to (d) refer to Fig.~\ref{fig:tp_eb10} %
and denote the configurations for which we studied the shape of the thermodynamic potential itself.}
\end{figure}

We can find the favorability of the \LOg~phase compared to the BCS and FF phases by finding the absolute minima of the thermodynamic potential. 
In Fig.~\ref{fig:tp_dm} we plot the local minima of the thermodynamic potential %
of three different ansatze, BCS, FF, and \LOg, for two binding energies, $\varepsilon_B=0.10\varepsilon_F$ (2) and $\varepsilon_B=0.25\varepsilon_F$ (1). Varying the dimensionless chemical potential difference, $\delta\mu/\Delta_0$, where $\Delta_0$ is the BCS order parameter at zero temperature, we determine the region where the \LOg~phase is favorable. The intersection of different phases determines the critical values  $\delta\mu_1$ (i.e., from the BCS to \LOg~transition) and $\delta\mu_2$ (i.e., from the \LOg~to \NPP~transition). 

\begin{figure}
\includegraphics[width=0.23\textwidth]{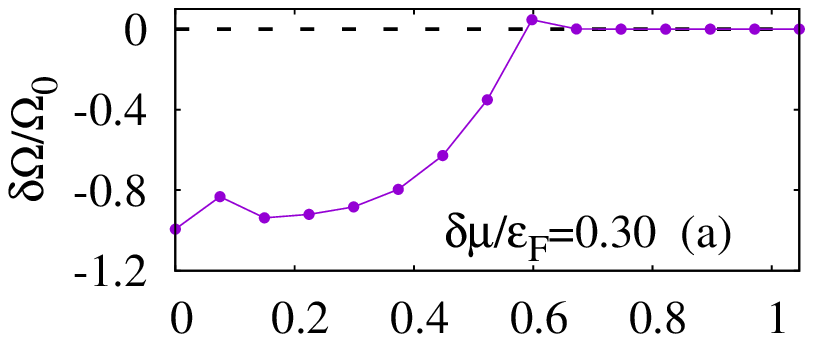}
\includegraphics[width=0.23\textwidth]{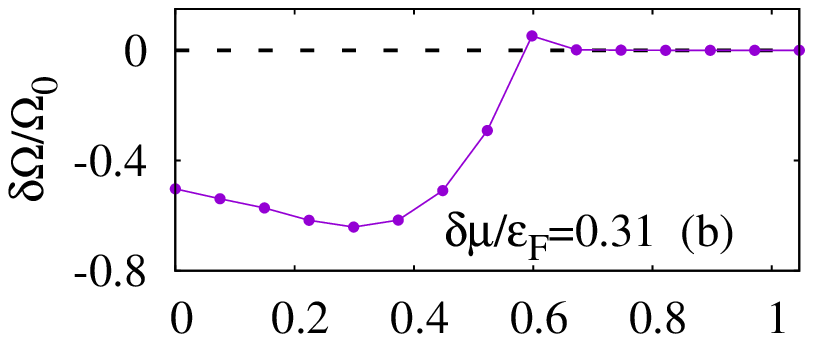}
\includegraphics[width=0.23\textwidth]{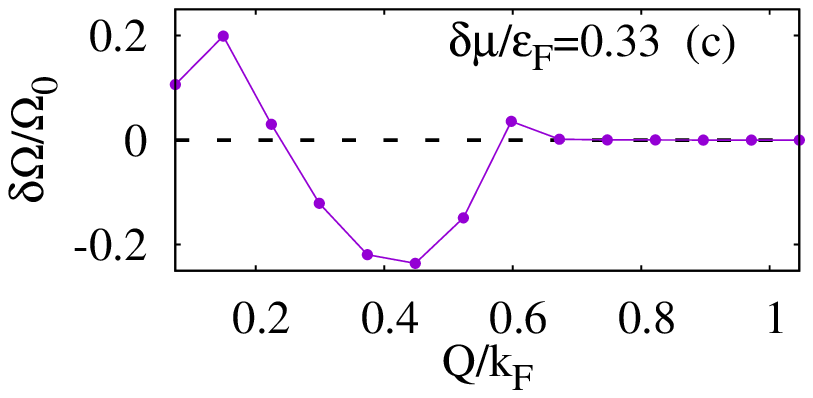}
\includegraphics[width=0.23\textwidth]{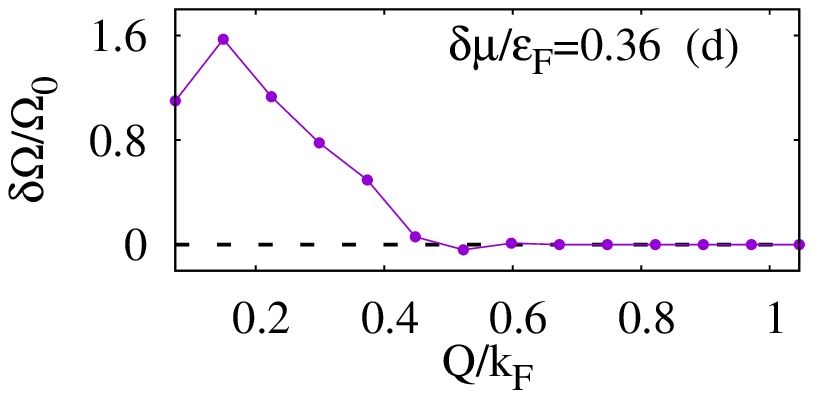}
\includegraphics[width=0.465\textwidth]{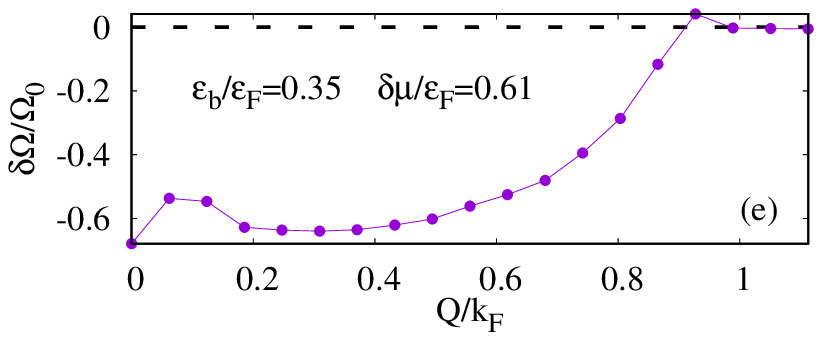}
\caption{\label{fig:tp_eb10}(color online). 
The thermodynamic potential as a function of $Q$ for interaction strength $\varepsilon_B = 0.1\varepsilon_F$ and chemical potential difference $\delta\mu=0.3$ (a),  $\delta\mu=0.31$ (b),  $\delta\mu=0.33$ (c), and  $\delta\mu=0.36$ (d). The plots (a)-(d) are labelled in Fig. 8. 
(e) BCS-\LOg~transition at %
higher binding energy $\epsilon_B=0.35\epsilon_F$ to emphasize the first-order nature of the transition. %
The thermodynamic potential is defined with respect to the free Fermi gas through $\delta\Omega=\Omega-\Omega_{\text{free}}$, %
and scaled by a constant, %
$\Omega_0=2\pi\times 10^{-3}N\epsilon_F$.
}
\end{figure}

We see in Fig.~\ref{fig:tp_dm} that the \LOg~to \NPP~transition is smooth \footnote{The FF line of minima is always above the \LOg~line %
close to the BCS line, although at high $\delta\mu$ values, the model accuracy doesn't allow us %
to conclude whether there is, or isn't, another transition FF-\LOg. Besides, we must say that is very %
inessential because, at high $\delta\mu$, the energy gap parameter reduces continuously its amplitude %
until it vanishes and the difference between the magnitude of the two ansatze might be neglegible. Moreover, %
at high $\delta\mu$ values it's even more likely, as discussed in Section~\ref{sec:fflo_family}, %
the presence of an LO$_2$ phase. The latter might actually be favourable with respect to both FF and \LOg.} %
while the BCS to \LOg~has a discontinuity, although %
not easy to observe. The order of the BCS-\LOg~transition can be examined as we increase the binding %
energy $\epsilon_B$. We check the behaviour of the order of phase transitions in Fig.~\ref{fig:tp_eb10}, where we have plotted the thermodynamic potential of the \LOg~phase as a function of $Q$ at the points labelled (a)-(d) in Fig.~\ref{fig:tp_dm} for the interaction strength $\epsilon_B=0.10\epsilon_F$, sliced at different $\delta\mu$ values. %
The comparison between subplots Fig.~\ref{fig:tp_eb10}(c) and (d) reveals the second order nature of the \LOg-\NPP~phase %
transition. In this part of the phase diagram, an increase of $\delta\mu$ leads to: %
(i) an increase of the LO momentum $Q$, (ii) the order parameter amplitude squeezes until it vanishes in $\delta\mu_2$ and (iii) the order %
parameter frequency matches the LO$_1$ initial ansatz. 
The BCS-\LOg~phase transition can be studied by observing the evolution of the thermodynamic %
potential in the subplots Fig.~\ref{fig:tp_eb10}(a) and (b). It is known that the FF phase undergoes a first order %
phase transition towards the BCS phase~\cite{Sheehy2015}. In our case, the presence of a local maximum at small $Q$ in the subplot Fig.~\ref{fig:tp_eb10}(a) is an indication of the first order phase transition. However, within our accuracy, the local maximum is shown by a single point only. It is possible %
to increase the binding energy, and thereby to relatively improve the accuracy of the calculation. As shown in Fig.~\ref{fig:tp_eb10}(e), at $\epsilon_B=0.35\epsilon_F$, the local maximum at low $Q$ becomes more evident. Considering the accuracy of our numerical calculations and also the difficulty of reaching convergence close to the BCS-\LOg~transition line, we cannot make a conclusive determination of the order of the BCS-\LOg~transition. Throughout the work, we interpret it as a \emph{weak} first-order transition, to reflect the fact that the strucuture of the local maxiumum is weak.

\subsection{Finite temperature}

\begin{figure}
\includegraphics[width=0.465\textwidth]{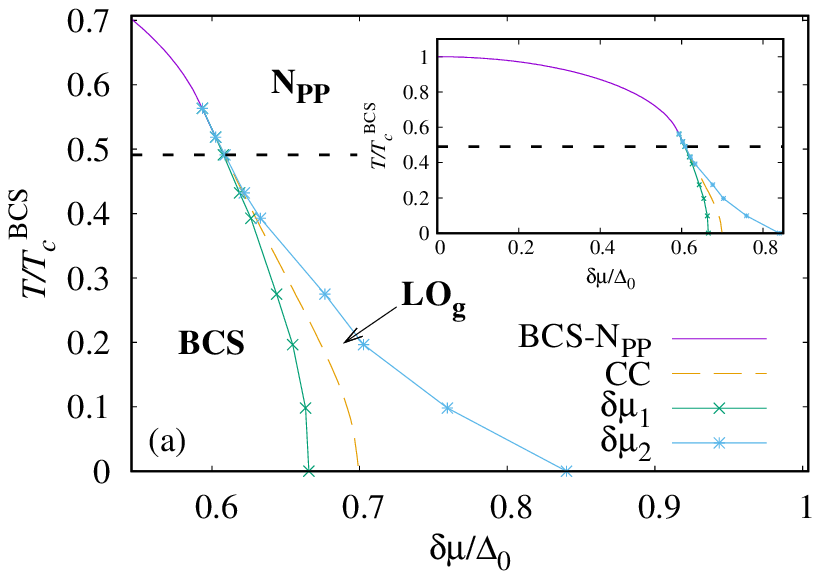}
\includegraphics[width=0.465\textwidth]{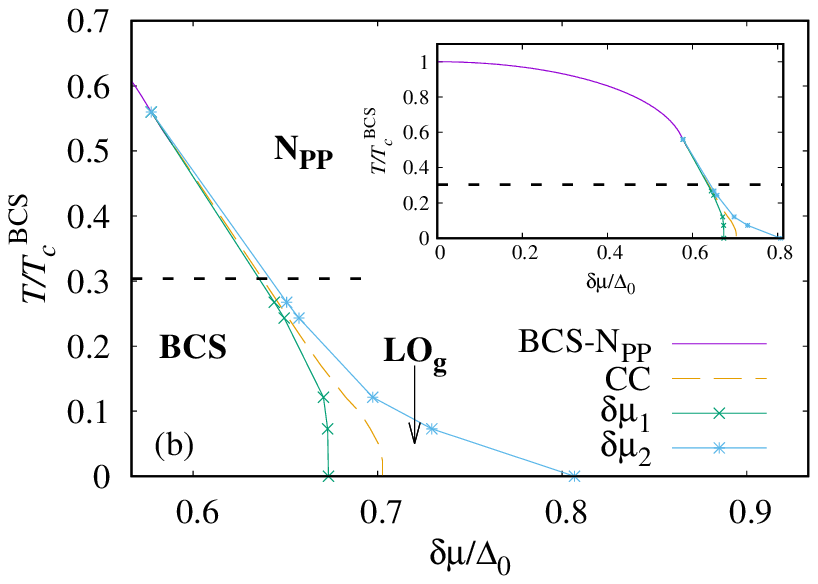}
\caption{\label{fig:t_vs_dm}(color online). 
Critical temperature of the second order BCS-\NPP~(solid) transition, the CC limit (dashed), weak first order BCS-\LOg~transition through $\delta\mu_1(T)$ (crosses), and second order \LOg-\NPP~through  $\delta\mu_2(T)$ (stars).  
The graph is scaled with respect to the BCS critical temperature $T_c^{\text{BCS}}$ at $\delta\mu=0$ and the BCS pairing gap at zero temperature $\Delta_0$. 
The inset shows the full phase diagram and the black dashed line is the BKT transition temperature.}
\end{figure}

We can extend the BdG equations from the zero temperature regime to study the \LOg~phase at finite temperatures. 
In the deep BCS limit the critical transition temperature, $T^{\text{BCS}}_c$, can be approximated %
from the gap equation in 2D following Landau theory of the second order phase transition, %
\be
T^{\text{BCS}}_c\simeq\frac{2}{\pi}e^{\gamma-\log(k_Fa_{\text{2D}})}T_F,
\ee 
where $\gamma$ is the Euler-Mascheroni constant. %
We can estimate the tricritical point (TCP) temperature, where the BCS, CC limit and FFLO phase intersect, %
with the numerical value \cite{Burkhardt1994,Matsuo1998,Combescot2002}
\be
T_{\text{TCP}}\simeq 0.561T^{\text{BCS}}_c.
\ee 
From our self-consistent BdG equations we plot out the phase diagram for the critical temperature as a function of chemical potential difference in Fig.~\ref{fig:t_vs_dm} for interaction strengths (a) $\epsilon_B=0.10\epsilon_F$ and (b) $\epsilon_B=0.25\epsilon_F$, the inset shows the entire phase diagram. The  transition between the BCS and \NPP~phase, found through the BCS theory, is second-order for temperatures above %
$T_{\text{TCP}}$. The  first-order CC limit (dashed line) is entirely contained %
in between the two \LOg~transition temperatures at any relevant $\delta\mu$ value. Moreover, the order of %
the phase transitions is preserved, 
the second order phase transition between the \LOg~and \NPP~phase is unaffected by an increasing temperature below TCP. %
The weak first-order BCS-\LOg~is preserved within our numerical accuracy. 

We note that, in the weakly-interacting regime $\epsilon_B=0.10\epsilon_F$, $T_c^{\text{BCS}}$ is approximately $0.25T_F$. It appears that the mean field treatment requires a beyond mean-field approach to take into %
account the fact there is no superfluid behaviour above the Berezinskii-€"Kosterlitz-€"Thouless (BKT) temperature $T_{\text{BKT}}\simeq0.125T_F$. %
We observe that most of the \LOg~phase is below $T_{\rm BKT}$, and within our numerical calculations the existence and availability of the \LOg~phase at finite temperatures is also below the TCP temperature, at least in the weakly-interacting regime. However, we cannot provide accurate quantitative results for the transition 
temperatures among the phases under investigation due to the mean-field nature of the calculation.

\section{\label{sec:sf}Superfluid density of the \LOg~phase}

The superfluid density is an essential signature of a superfluid system, and as we have seen the structure of the pairing order parameter has been significantly altered for a finite polarization. We can study the superfluid density of a Fermi gas by imposing a superfluid twist~\cite{Taylor2006,PhysRevA.75.033609,PhysRevB.77.144521,PhysRevB.76.092502}.
The superfluid part of the system, if it exists, cannot react to small excitations like a normal gas and creates a superfluid current, whose magnitude is proportional to the superfluid density contained in the gas~\cite{Fisher1973}. We impose the twist by modifying the pairing order parameter in Eq.~\eqref{eq:eigenprob2},  
\be\label{eq:twist}
\Delta(\vect{x}) = \Delta(\vect{x})e^{\im \vect{Q}_s\cdot \vect{x}},
\ee
where the twist momentum is $\vect{Q}_s$ and the superfluid velocity $\vect{v}_s$ is given by, 
\be
\vect{v}_s=\frac{\vect{Q}_s}{2m}.
\ee
The Helmholtz free energy, $F$, must increase if we impose a flow in the superfluid phase and %
the shape of the free energy for small twists is shown to behave like the square of the phase twist %
amplitude around the point $F(\vect{Q}_s=\vect{0})$~\cite{Taylor2006},  
\begin{alignat}{1}
\label{eq:dF}
\frac{F(\vect{Q}_s)-F(\vect{0})}{V} & \simeq  %
\frac{|\vect{Q}_s|^2}{2V}\left(\frac{\partial^2F(\vect{Q}_s)}{\partial|\vect{Q}_s|^2}\right)_{\vect{Q}_s\rightarrow\vect{0}} \equiv \frac{1}{2}\rho_smv_s^2,
\end{alignat}
where the superfluid density $\rho_s$ is, 
\be\label{eq:superfluid_density}
\rho_s=4m\left(\frac{\partial^2F(\vect{Q}_s)}{\partial|\vect{Q}_s|^2}\right)_{\vect{Q}_s=\vect{0}},
\ee
and the normal part density must be $\rho_n=n-\rho_s$.

\begin{figure}
\includegraphics[width=0.48\textwidth]{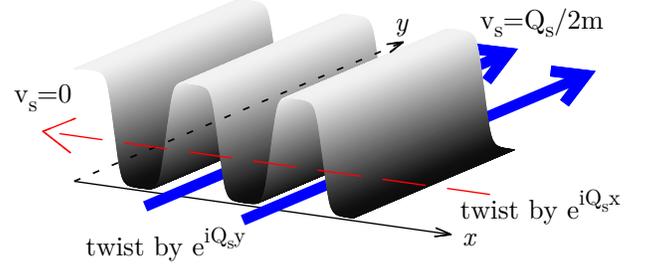}
\caption{\label{fig:sketch}(color online). Effect of a phase twist imposed on the pairing order parameter. In the parallel %
direction (red broken line), superfluid current can hardly be produced, while in the perpendicular direction (blue line), superfluid current is allowed to flow, except the regions where the order parameter vanishes and the spin imbalance is high.}
\end{figure}

Due to the symmetry we have imposed on the \LOg~order parameter and from Fig.~\ref{fig:delta_pola}, %
where the local spin polarization is peaked when $\Delta(\mathbf{x})$ vanishes, we can infer the fundamental properties of the superfluid phase. 
In Fig.~\ref{fig:sketch}, we illustrate the idea of what might happen when we
impose a superfluid twist that is either parallel ($x$-direction) or perpendicular ($y$-direction) to the LO order parameter. 
A macroscopic current along the $x$-direction must cross regions %
where the \LOg~phase has domain walls and thus cannot flow. %
A superfluid current along the $y$-direction is able to flow as the order parameter is nonzero. 
By setting $\vect{Q}_s=Q_s\hat{\vect{x}}$, we find that there is no detectable superfluid density within our numerical precision. Thus, we restrict the following analysis to the perpendicular case, choosing $\vect{Q}_s=Q_s\hat{\vect{y}}$, and computing the superfluid density via Eq.~\eqref{eq:superfluid_density} through a modification of the BdG equations given in Section~\ref{sec:model}.

Equations~\eqref{Eq:bdgfinal} in the presence of the twist are now complex valued and need to be re-cast in a real-valued form. In order to achieve this we have to consider again the Heisenberg equations of motion and perform a new Bogoliubov transformation including the twist order parameter Eq.~\eqref{eq:twist}. The new modes will be decomposed as follows,
\begin{alignat}{1}
\psi_\uparrow(\vect{x})= & %
\sum_\eta\biggl[u_{\eta \vect{Q}_s\uparrow}(\vect{x})c_{\eta \vect{Q}_s\uparrow}e^{-\im E_{\eta \vect{Q}_s\uparrow} t} \nonumber \\
&\qquad\qquad-%
v^*_{\eta \vect{Q}_s\downarrow}(\vect{x})c^{\dagger}_{\eta\vect{Q}_s\downarrow}e^{\im E_{\eta\vect{Q}_s\downarrow} t}\biggl], \nonumber \\
\psi^{\dagger}_\downarrow(\vect{x}) = & %
\sum_\eta\biggl[u^*_{\eta\vect{Q}_s\downarrow}(\vect{x})c^{\dagger}_{\eta\vect{Q}_s\downarrow}e^{\im E_{\eta\vect{Q}_s\downarrow} t} \nonumber \\
&\qquad\qquad +%
v_{\eta\vect{Q}_s\uparrow}(\vect{x})c_{\eta\vect{Q}_s\uparrow}e^{-\im E_{\eta\vect{Q}_s\uparrow} t}\biggl],
\end{alignat}
yielding 
\be\begin{split}
&M_\uparrow\twovec{u_{\eta\vect{Q}_s\uparrow}}{v_{\eta\vect{Q}_s\uparrow}}=\\
&=\fourvec{-\nabla^2_{\vect{x}}-\mu_\uparrow}%
{-\Delta(\vect{x})e^{\im \vect{Q}_s\cdot\vect{x}}}{-\Delta(\vect{x})^*e^{-\im \vect{Q}_s\cdot\vect{x}}}%
{\nabla^2_{\vect{x}}+\mu_\downarrow}\twovec{u_{\eta \vect{Q}_s\uparrow}}{v_{\eta \vect{Q}_s\uparrow}}.
\end{split}\ee
We can remove the phase from the the order parameter through the transformation $AM_\uparrow A^*$, where
\be
A=
\begin{bmatrix}
e^{\frac{\im}{2} \vect{Q}_s\cdot\vect{x}} & 0 \\
0 & e^{-\frac{\im}{2} \vect{Q}_s\cdot\vect{x}}
\end{bmatrix}.
\ee
As in the derivation of the FFLO BdG equations, the two spin components are related through a unitary transformation and we only need to study one spin component. %
We thus define $u_{\eta \vect{Q}_s\uparrow}=u_{\eta \vect{Q}_s}$ and $v_{\eta \vect{Q}_s\uparrow}=v_{\eta \vect{Q}_s}$, obtaining the following form for the BdG equations, %
\be\begin{split}\label{eq:superf_bdg}
&\fourvec{
-\left(\vec{\nabla}+\im\frac{\vect{Q}_s}{2}\right)^2-\mu_\uparrow}{-\Delta(\vec{x})}{-\conj{\Delta(\vec{x})}}{%
\left(\vec{\nabla}-\im\frac{\vect{Q}_s}{2}\right)^2+\mu_\downarrow}\twovec{u_{\eta \vect{Q}_s}}{v_{\eta \vect{Q}_s}}=\\
&=E_{\eta}^{\vect{Q}_s}%
\twovec{u_{\eta \vect{Q}_s}}{v_{\eta\vect{Q}_s}}.
\end{split}\ee
Choosing a %
Hilbert basis of complex-valued functions,
$$
\phi_{\vect{k}}(\vect{x})=\frac{e^{\im\vect{k}\cdot\vect{x}}}{\sqrt{V}},\qquad \vect{k}=\frac{2\pi}{\sqrt{V}}\vect{n},
$$
with $\vect{n}\in\ZZ^2$, 
we expand the quasi-particle wavefunctions as,  
$$
u_{\eta \vect{Q}_s}=\sum_{\vect{k}}A^{(\eta \vect{Q}_s)}_{\vect{k}}\phi_{\vect{k}},\qquad 
v_{\eta \vect{Q}_s}=\sum_{\vect{k}}B^{(\eta \vect{Q}_s)}_{\vect{k}}\phi_{\vect{k}},
$$
and expand the BdG equations with the requirement that $\Delta(\vect{x})$ is real-valued, uni-directional function %
and a superposition of cosine Fourier modes. As in Section~\ref{sec:model}, we obtain a block-diagonal set of matrix equations in $n_2$, giving
\begin{alignat}{1} 
\fourvec{\epsilon^{\uparrow}_{nm}}{-\Delta_{n_1m_1}}{-\Delta_{nm}}{-\epsilon^{\downarrow}_{nm}}
\twovec{A^{(\eta \vect{Q}_s n_2)}_{m} }{B^{(\eta  \vect{Q}_s n_2)}_{m}} 
=E_{\eta}^{\vect{Q}_s n_2}
\twovec{A^{(\eta \vect{Q}_s n_2)}_{n} }{B^{(\eta \vect{Q}_s n_2)}_{n}}, 
\end{alignat}
where we define the two single particle energies,
\begin{alignat}{1}
\epsilon^{\uparrow}_{nm} &=
\left[\left(\vect{k}+\frac{\vect{Q}_s}{2}\right)^2-\mu_\uparrow\right] \delta_{nm}, \nonumber \\
\epsilon^{\downarrow}_{nm} &=
\left[\left(\vect{k}-\frac{\vect{Q}_s}{2}\right)^2-\mu_\downarrow\right] \delta_{nm}.
\end{alignat}
The order parameter matrix elements are found from,
\be
\Delta_{nm}=\frac{1}{L}\int_{-L/2}^{L/2} dx\ \cos((n-m)x)\Delta(x),
\ee
The choice of the cut-off energy is matched %
to the $\vect{Q}_s=0$ calculations and adjusted to ensure the calculation is independent from the cut-off choice. 
\begin{figure}
\includegraphics[width=0.48\textwidth]{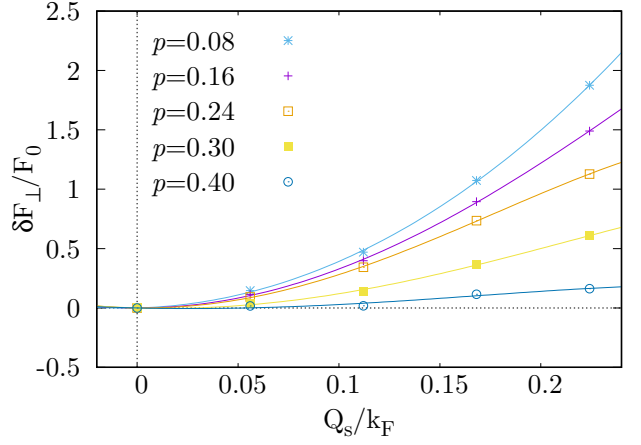}
\caption{\label{fig:sf}(color online). The perpendicular %
Helmholtz free energy difference, $\delta F_\bot=F(\vect{Q}_\bot)-F(\vect{0})$, as a function of the phase twist, %
for various polarizations. The free energy variation is calculated by using $\vect{Q}_{\bot}=Q_s\hat{y}$ and is plotted %
in units of a rescaling constant $F_0=2\pi\times 10^{-3}N\epsilon_F$. %
The continuous lines are %
the data fit using a two-parameter polynomial curve $ax^2+bx^4$. 
}
\end{figure}
\begin{figure}
\includegraphics[width=0.48\textwidth]{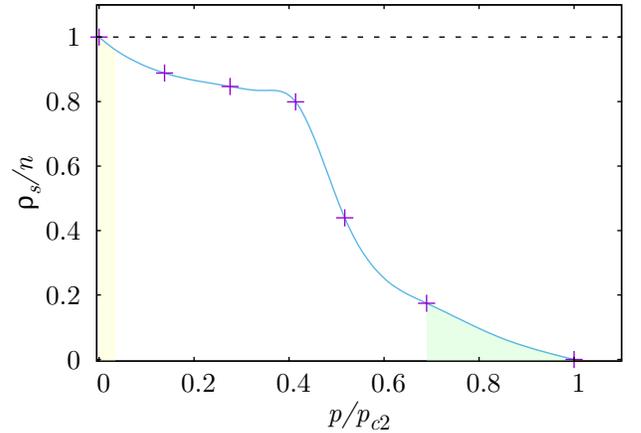}
\caption{\label{fig:d_vs_p}(color online). The superfluid fraction $\rho_s/n$ as a function of the polarization $p$ at $\epsilon_B=0.25\epsilon_F$ at zero %
temperature. The polarization is rescaled by the critical polarization $p_{c2}=0.58$, at which the \LOg-\NPP~transition occurs. The green and yellow regions %
correspond to the \LOg$\rightarrow$LO$_1$ and BCS regions shown in FIG.~\ref{fig:dm_p}, respectively. The solid line %
is a guide line.}
\end{figure}

In Fig.~\ref{fig:sf} we show the Helmholtz free energy under a phase twist $\vect{Q}_s=\vect{Q}_\bot=Q_s\hat{y}$ with increasing twist amplitude (measured in units of the Fermi momentum, $k_F$) at zero temperature and interaction strength $\varepsilon_B=0.25\varepsilon_F$. 
The parabolic dependency in $Q_s$ is obtained by fitting the data with polynomial cure, $ax^2+bx^4$. We find the coefficient $b$ is negligible and the coefficient $a$ is then proportional to the superfluid density of the system using Eq.~\eqref{eq:superfluid_density}.  
We note that, in the presence of the LO order parameter, the calculation of the superfluid density becomes very difficult. The only way to increase the precision of our approach is to increase the size of the system, i.e. $L$. This can be done by increasing the number of particles, however, this greatly increases the computational cost. Previous computations related to the FF phase~\cite{Cao2015}, which use similar methods, have been able to dramatically increase the accuracy of numerical calculations,  approximately 10 times better than our accuracy.

In Fig.~\ref{fig:d_vs_p}, we present the superfluid fraction as a function of the polarization at the interaction strength $\epsilon_B=0.25\epsilon_F$ and at zero temperature. In general, we observe that an increase of polarization suppresses the superfluid density. 
As the LO momentum of the pairing gap increases with increasing polarization,
as shown in Fig. 3(c), the local spin imbalance becomes more spread in space, making it more difficult for the superfluid current to flow.
We note that there is a significant drop in the superfluid density for polarization larger than $p\simeq0.4p_{c2}$. 
This behavior may be due to the high polarization causing the pairing order parameter to be highly oscillatory, where the \LOg~and LO$_1$ phases are matching, increasing the number of zeros in the order parameter and preventing the superfluid current from flowing. 
As a result, the superfluid fraction decreases dramatically, making this regime difficult to experimentally probe.

\section{\label{sec:conclusions}Conclusions}

We have studied in detail the mean-field theory of a two-component 2D atomic Fermi gas in the presence of spin-population imbalance.  The pairing order parameter has been self-consistently determined through numerical solutions of the BdG equations in the case of an $s$-wave scattering contact potential, allowed only between opposite spin particles. We have considered different types of pairing order parameter and explored the phase transitions and the superfluid nature of the system. In particular, the LO ansatz for the pairing order parameter has been refined and treated with reasonable approximations in order to make computations accessible and useful. In constrast to the previous studies with a FF ansatz~\cite{Conduit2008,Sheehy2015}, such a refinement reveals the pivotal importance of the \LOg~ansatz among the broad FFLO family in 2D. The \LOg~phase turns out to be energetically favourable with respect to both the FF and the original LO ansatz. The superfluid density of the \LOg~phase has been calculated by adding a phase twist to the pairing order parameter. We have predicted the qualitative behavior of the superfluid density with increasing spin polarization.

The mathematical and computational methods employed in our work are satisfactory, although they are incomplete for providing exact quantitative results. Many of the configurations under our investigation mimic some experimental settings for 2D or quasi-2D Fermi gases~\cite{Dyke2016}. Therefore, we anticipate that our results might be useful for possible experimental realizations of a FFLO phase in 2D in the near future. We note also that a beyond-mean-field treatment is required to obtain quantitative results and to estimate the critical temperature of the \LOg~phase. 

\begin{acknowledgments}
We would like to thank Chris Vale for stimulating discussions and S. Kiesewetter for his help and advice during the early stage of this work. This work was supported by the ARC Discovery Projects: DP140100637 and FT140100003 (XJL), FT130100815 and DP140103231 (HH).
\end{acknowledgments}

\bibliographystyle{apsrev4-1}
\bibliography{main_bib_2}

\end{document}